\documentclass[aps,prd,groupedaddress,amsmath,reprint,nofootinbib,twocolumn,10pt]{revtex4-1}

\usepackage{color,hhline,psfrag,rotating,amssymb}
\usepackage{dcolumn}
\usepackage{verbatim}
\usepackage{hyperref}
\usepackage[all]{hypcap}

\newcommand{\be}{\begin{equation}}
\newcommand{\ee}{\end{equation}}
\newcommand{\bea}{\begin{eqnarray}}
\newcommand{\eea}{\end{eqnarray}}

\begin{document}
\title{$B$-meson decay constants: a more complete picture from full 
lattice QCD}

\author{B.~Colquhoun}
\affiliation{SUPA, School of Physics and Astronomy, University of Glasgow, Glasgow, G12 8QQ, UK}
\author{C.~T.~H.~Davies}
\email[]{christine.davies@glasgow.ac.uk}
\affiliation{SUPA, School of Physics and Astronomy, University of Glasgow, Glasgow, G12 8QQ, UK}
\author{R.~J.~Dowdall}
\affiliation{DAMTP, University of Cambridge, Wilberforce Road, Cambridge, CB3 0WA, UK}
\author{J.~Kettle}
\affiliation{SUPA, School of Physics and Astronomy, University of Glasgow, Glasgow, G12 8QQ, UK}
\author{J.~Koponen}
\affiliation{SUPA, School of Physics and Astronomy, University of Glasgow, Glasgow, G12 8QQ, UK}
\author{G.~P.~Lepage}
\affiliation{Laboratory of Elementary-Particle Physics, Cornell University, Ithaca, NY 14853, USA}
\author{A.~T.~Lytle}
\affiliation{SUPA, School of Physics and Astronomy, University of Glasgow, Glasgow, G12 8QQ, UK}
\collaboration{HPQCD collaboration}
\homepage{http://www.physics.gla.ac.uk/HPQCD}
\noaffiliation

\date{\today}

\begin{abstract}
We extend the picture of $B$-meson decay constants obtained in lattice QCD 
beyond those of the $B$, $B_s$ and $B_c$ to give the first 
full lattice QCD results for the $B^*$, $B^*_s$ and $B^*_c$.  
We use improved NonRelativistic QCD for the valence $b$ quark 
and the Highly 
Improved Staggered Quark (HISQ) action for the lighter quarks on gluon field 
configurations that include the effect of $u/d$, $s$ and $c$ quarks in the 
sea with $u/d$ quark masses going down to physical values. 
For the ratio of vector to pseudoscalar decay constants, 
we find $f_{B^*}/f_B$ = 0.941(26), $f_{B^*_s}/f_{B_s}$ = 0.953(23) 
(both $2\sigma$ less than 1.0) 
and $f_{B^*_c}/f_{B_c}$ = 0.988(27). Taking 
correlated uncertainties into account we see clear indications 
that the ratio increases as the 
mass of the lighter quark increases.  
We compare our results to those using the HISQ formalism for all 
quarks and find good agreement both on decay 
constant values when the heaviest quark is a $b$ 
and on the dependence on the mass of the heaviest quark in the region 
of the $b$. 
Finally, we give an overview plot of decay constants for gold-plated 
mesons, the most complete picture of these hadronic parameters to date.

\end{abstract}


\maketitle

%
\section{Introduction}
\label{sec:intro}
%
Lattice QCD calculations are now an essential part of $B$ physics 
phenomenology (see for example~\cite{Davies:2012qf}), providing increasingly 
precise determinations of decay constants, form factors 
and mixing parameters needed, 
along with experiment, in the determination of Cabibbo-Kobayashi-Maskawa 
(CKM) matrix elements. 
As the constraints being provided by lattice QCD 
become more stringent it is increasingly important to expand 
the range of hadronic matrix elements being calculated to allow tests both 
against experiment where possible and/or against expectations from 
other approaches. Decay constants are particularly useful in this 
respect because they are single numbers expressing the amplitude for a 
meson to annihilate to a single particle (for example a 
$W$ boson or a photon), encapsulating information about its internal 
structure. They are straightforwardly calculated in 
lattice QCD from the same hadron correlation functions being 
used to determine the hadron masses. The only additional 
complication is that 
normalisation of the appropriate operator for the meson 
creation/annihilation is required. 
In this way we can build up a tested and 
consistent `big picture' 
of meson decay constants within which sit the results 
being used for CKM element determination. 

To this end we determine here the decay constants that parameterise 
the amplitude to annihilate for the vector 
mesons $B^*$ and $B_s^*$. These mesons are the partners of the $B$ and 
$B_s$ whose weak decay matrix elements are 
critical to understanding heavy flavour 
physics. Decay modes of the $B^*$ and $B_s^*$ are dominated 
by electromagnetic radiative decays~\cite{pdg} to $B$ and $B_s$, 
however, and 
so it is unlikely that processes 
in which the decay constant is the key hadronic parameter will be measured 
experimentally. The determination of the vector decay constants is 
nevertheless useful 
because the relationship with that of the pseudoscalar decay constant 
can be understood within the framework of Heavy Quark Effective Theory (HQET) 
and the decay constants appear in phenomenological analyses of 
the vector form factor for semileptonic decay processes for the pseudoscalar 
mesons (see~\cite{Becirevic:2014kaa} for a recent discussion of this). 

Since vector and pseudoscalar heavy-light mesons differ only 
in their internal spin configuration, their decay constants might 
be expected to have rather similar values. The key question is then: by 
how much do they differ and which is larger? 
A recent review~\cite{Lucha:2014hqa} showed tension between
the results for the ratio of the $B^*$ to $B$ decay constants from 
QCD sum rules and from lattice QCD. The lattice QCD results
 used $u/d$ quarks (only) in the sea and obtained results for mesons 
containing $b$ quarks from an 
interpolation between results for quarks close to the $c$ mass and the static 
(infinite mass) limit~\cite{Becirevic:2014kaa}. This gave a result 
for the ratio greater than 1 whereas the QCD sum rules approach quoted 
preferred a value less than 1. 

The results we give here build on our state-of-the-art calculation of 
the $B$ and $B_s$ decay constants~\cite{Dowdall:2013tga} 
using an improved NonRelativistic QCD 
(NRQCD) formulation~\cite{Dowdall:2011wh} that allows us 
to work to high accuracy directly at 
the $b$ quark mass. We also use lattice QCD gluon field configurations that 
have the most realistic QCD vacuum to date,
include $u/d$, $s$ and $c$ quarks in the sea (using the 
Highly Improved Staggered Quark formalism~\cite{Follana:2006rc}) 
with the $u/d$ quark 
mass taking values down to the physical value. We therefore avoid
significant systematic errors from extrapolations in the $u/d$ quark mass.  
We are able to give results for the ratio of vector to 
pseudoscalar decay constants for both the $B_s$ and the $B$ 
and the SU(3)-breaking ratio of these ratios. 
We find clearly that the vector decay constant 
is smaller than the pseudoscalar decay constant in both cases. 

We also give results for the decay constant of the $B_c$ meson and 
its vector partner the $B_c^*$. The $B_c$ has been seen experimentally 
only relatively recently~\cite{pdg} and is interesting 
because it can be viewed both 
as a heavy-heavy meson and as a heavy-light meson. 
Here we compare its ratio of vector to pseudoscalar decay 
constants to that of the $B$ and $B_s$, and find the ratio 
is significantly larger, now being very close to 1. 

The decay constant of the $B_c$ is quite different from 
that of the $B$ and $B_s$, being nearly double their size. 
The $B_c$ decay constant can be used to predict its partial 
width for leptonic decay that may be observed in the future. 
We determine this decay constant here using NRQCD 
$b$ quarks and HISQ $c$ quarks and compare to our 
previous result~\cite{McNeile:2012qf} 
that used the HISQ action for both quarks and  
mapped out the behaviour of a range of decay constants 
for valence heavy quarks in the region between the $c$ quark 
mass and the $b$ quark mass. Since the HISQ action is fully 
relativistic this is a good test of our understanding of systematic 
errors in lattice QCD, and confirmation of how well improved 
actions work. 

In a further study of this point we go on to look at the 
dependence of decay constants on the valence heavy 
quark mass using quark masses 
lighter than that of the $b$ in the NRQCD action. This 
enables us to compare both the value of specific decay 
constants and the dependence on the heavy quark mass with 
that from using the relativistic HISQ action. 
We also demonstrate the consistency of our 
results for the ratio of vector to pseudoscalar decay 
constants for the $B_s$ meson here to our earlier result for 
the same ratio for the $D_s$ meson~\cite{Donald:2013pea}
using HISQ quarks. 

A very consistent picture thus emerges from both a 
nonrelativistic and a relativistic approach to heavy 
quarks within lattice QCD. Both approaches are the 
result of several stages of improvement
to reduce discretisation errors and other systematic 
uncertainties to a low level, important 
for making a detailed comparison. 

We begin by outlining the methods used in our lattice calculation, 
which follow~\cite{Dowdall:2013tga,Dowdall:2012ab}.
Section~\ref{subsec:BBs} gives results for the decay constants 
of the $B_s^*$ and $B^*$ 
and their comparison, and then section~\ref{subsec:Bc} gives 
results for the decay constants of the $B_c$ and the $B_c^*$. 
Section~\ref{subsec:mQ} works with quarks lighter than $b$ to 
demonstrate the heavy quark mass dependence of the decay constants and 
compare to our earlier results using the HISQ formalism for $b$ quarks. 
Section~\ref{sec:discussion} compares our results for vector 
meson decay constants to those of earlier 
determinations using other methods, including HQET arguments, and 
shows how the $B_c$ fits in between results for heavyonium and 
heavy-light mesons. 
Section~\ref{sec:conclusions} gives our 
conclusions, including the promised `big picture' for the 
decay constants of gold-plated mesons from lattice QCD, the most complete 
picture of these hadronic parameters to date. 
%
\section{Lattice calculation}
\label{sec:lattice}
Since the first lattice NRQCD calculations were done for 
heavy-light mesons~\cite{Davies:1993qp}, huge improvements have been made. The 
current state-of-the-art~\cite{Dowdall:2013tga, Dowdall:2012ab} 
uses an improved NRQCD action for the heavy quark coupled to 
a HISQ light quark on gluon configurations that include an improved 
gluon action and HISQ sea quarks. Here we extend these calculations 
to include the decay constants of vector heavy-light mesons.  

The gluon field configurations that we use were generated by 
the MILC collaboration~\cite{Bazavov:2010ru, Bazavov:2012xda}.  
These are $n_f = 2+1+1$ configurations that include the 
effect of light (up/down), strange and charm quarks in the sea with the HISQ 
action~\cite{Follana:2006rc, oldfds} 
and a Symanzik improved gluon action with coefficients correct 
through $\mathcal{O}(\alpha_s a^2,n_f\alpha_s a^2)$ \cite{Hart:2008sq}. 
The lattice spacing 
values that we use range from $a=0.15$ fm to $a=0.09$ fm. 
The configurations have well-tuned 
sea strange quark masses and sea light quark masses ($m_u=m_d=m_l$) 
with ratios to the strange 
mass from $m_l/m_s=0.2$ down to the value that corresponds to the 
experimental $\pi$ meson mass of $m_l/m_s=1/27.4$~\cite{Bazavov:2014wgs}. 

In~\cite{Dowdall:2011wh} we accurately determined the 
lattice spacings using the mass difference of the 
$\Upsilon^{\prime}$ and $\Upsilon$ mesons using the same 
NRQCD action for the $b$ quark as we use here. 
The details of each ensemble, including the lattice spacing, sea quark masses 
and spatial volumes, are given in table \ref{tab:params}. 
All ensembles were fixed to Coulomb gauge.

\begin{table*}
\caption{
Details of the ensembles (sets) of gauge field configurations 
used in this calculation \cite{Bazavov:2010ru, Bazavov:2012xda}. 
$\beta$ is the bare gauge coupling,
$a_{\Upsilon}$ is the lattice spacing as determined by the $\Upsilon(2S-1S)$ splitting in \cite{Dowdall:2011wh}, where the three errors are statistics, NRQCD systematics and experiment. 
Column 4 gives the corresponding values of $\alpha_s$ used in 
renormalisation factors. This is taken as  
$\alpha_V(n_f=4,q=2/a)$ and determined 
from~\cite{McNeile:2010ji, Chakraborty:2014aca}. 
$am_l,am_s$ and $am_c$ are the sea quark masses in lattice units. 
We also give, in column 8, values for $\delta x_{\mathrm{sea}}$, the 
fractional difference in the sum of the light quark masses from 
their physical values. 
$\delta x_{\mathrm{sea}}$ is defined as 
$(2m_l+m_s)/(2m_{l,\mathrm{phys}}+m_{s,\mathrm{phys}})-1$, 
using values of $m_{s,\mathrm{phys}}$ from~\cite{Dowdall:2011wh} 
and $m_s/m_l=27.4$~\cite{Bazavov:2014wgs}. 
$L/a \times T/a$ gives the spatial and temporal extent of the 
lattices and $n_{{\rm cfg}}$ is the number of configurations in each ensemble. 
16 time sources were used for the valence quark propagators 
on each configuration for increased 
statistics.  
Sets 1, 2 and 3 will be referred to in the text as ``very coarse'', 
4, 5 and 6 as ``coarse'' and 7 as ``fine''. Sets 3 and 6 include light sea 
quarks with their physical masses. 
}
\label{tab:params}
\begin{ruledtabular}
\begin{tabular}{lllllllllll}
Set & $\beta$ & $a_{\Upsilon}$ (fm) & $\alpha_V(2/a)$	& $am_{l}$ & $am_{s}$ & $am_c$ & $\delta x_{\mathrm{sea}}$ & $L/a \times T/a$ & $n_{{\rm cfg}}$  \\
\hline
1 & 5.80 & 0.1474(5)(14)(2) & 0.346 & 0.013   & 0.065  & 0.838 & 0.323 & 16$\times$48 & 1020 \\
2 & 5.80  & 0.1463(3)(14)(2)& 0.344 & 0.0064  & 0.064  & 0.828 & 0.126 & 24$\times$48 & 1000 \\
3 & 5.80  & 0.1450(3)(14)(2)& 0.343  & 0.00235  & 0.0647  & 0.831 & 0.027 & 32$\times$48 & 1000 \\
\hline
4 & 6.00 & 0.1219(2)(9)(2) & 0.311 & 0.0102  & 0.0509 & 0.635 & 0.259 & 24$\times$64 & 1052 \\
5 & 6.00 & 0.1195(3)(9)(2) & 0.308 & 0.00507 & 0.0507 & 0.628 & 0.108 & 32$\times$64 & 1000 \\
6 & 6.00 & 0.1189(2)(9)(2) & 0.307 & 0.00184 & 0.0507 & 0.628 & -0.004 & 48$\times$64 & 1000 \\
\hline
7 & 6.30 & 0.0884(3)(5)(1) & 0.267 & 0.0074  & 0.0370  & 0.440 & 0.327 & 32$\times$96 & 1008 \\
\end{tabular}
\end{ruledtabular}
\end{table*}
%
%
%
\subsection{NRQCD valence quarks}
\label{subsec:nrqcd}
%
We use improved NRQCD for the $b$ quark, which takes 
advantage of the nonrelativistic nature of the $b$ quark within 
its bound states for very good control of discretisation uncertainties.
This allows us to work with relatively low numerical cost
on the lattices with the lattice spacing 
values given above.
NRQCD has the advantage 
that the same action can be used for both bottomonium and 
$B$-meson calculations so that tuning of the $b$-quark mass and 
determination of the lattice spacing can be 
done using bottomonium and there are no new parameters to be tuned at 
all for $B$-mesons. $b$-quark propagators are calculated in NRQCD by 
evolving forward in time (using eq.~(\ref{eq:Hamiltonian})) 
from a starting condition. 
This is numerically very fast and
high statistics can then readily be accumulated for precise results. 
The action used here builds on the standard NRQCD 
action~\cite{Lepage:1992tx} accurate through $v^4$ 
in the heavy quark velocity $v$ 
(using power-counting terminology for bottomonium) by 
including one loop radiative corrections to many of the 
$v^4$ coefficients~\cite{Hammant:2011bt, Dowdall:2011wh}.
We studied the effect of these improvements on the bottomonium 
spectrum in~\cite{Dowdall:2011wh, Daldrop:2011aa, Dowdallhyp} and in $B$, $B_s$ and $B_c$ meson 
masses in~\cite{Dowdall:2012ab}. 

The NRQCD Hamiltonian we use is given by \cite{Lepage:1992tx}:
 \begin{eqnarray}
\label{eq:evolution}
 e^{-aH} &=& \left(1-\frac{a\delta H}{2}\right)\left(1-\frac{aH_0}{2n}\right)^n U_t^{\dag} \nonumber \\
&& \times \left(1-\frac{aH_0}{2n}\right)^n\left(1-\frac{a\delta H}{2}\right) 
\end{eqnarray}
with
 \begin{eqnarray}
 aH_0 &=& - \frac{\Delta^{(2)}}{2 am_b}, \nonumber \\
a\delta H
&=& - c_1 \frac{(\Delta^{(2)})^2}{8( am_b)^3}
            + c_2 \frac{i}{8(am_b)^2}\left(\bf{\nabla}\cdot\tilde{\bf{E}}\right. -
\left.\tilde{\bf{E}}\cdot\bf{\nabla}\right) \nonumber \\
& & - c_3 \frac{1}{8(am_b)^2} \bf{\sigma}\cdot\left(\tilde{\bf{\nabla}}\times\tilde{\bf{E}}\right. -
\left.\tilde{\bf{E}}\times\tilde{\bf{\nabla}}\right) \nonumber \\
 & & - c_4 \frac{1}{2 am_b}\,{\bf{\sigma}}\cdot\tilde{\bf{B}}  
  + c_5 \frac{\Delta^{(4)}}{24 am_b} \nonumber \\
 & & -  c_6 \frac{(\Delta^{(2)})^2}{16n(am_b)^2} .
\label{eq:Hamiltonian}
\end{eqnarray}
Here $\nabla$ is the symmetric lattice derivative and $\Delta^{(2)}$ and 
$\Delta^{(4)}$ the lattice discretization of the continuum $\sum_iD_i^2$ and 
$\sum_iD_i^4$ respectively. $am_b$ is the bare $b$ quark mass. 
The parameter $n$ has no physical significance, but 
is included for numerical stability of high momentum modes. We take the value 
$n=4$ here in all cases. 
$\bf \tilde{E}$ and $\bf \tilde{B}$ are the chromoelectric 
and chromomagnetic fields calculated from an improved clover term~\cite{Gray:2005ur}.
The $\bf \tilde{B}$ and $\bf \tilde{E}$ are made anti-hermitian 
but not explicitly traceless, to match the perturbative calculations 
done using this action.  

The coefficients $c_i$ in the action are unity at tree level but radiative corrections cause them to depend on $am_b$ at higher orders in 
$\alpha_s$. These were calculated for the relevant $b$ quark masses 
using lattice perturbation theory in~\cite{Dowdall:2011wh, Hammant:2011bt} 
and the 
values used in this paper are given in Table~\ref{tab:wilsonparams}. 
Including the one-loop radiative corrections 
to $c_4$ is particularly important here, 
since this coefficient controls the hyperfine splitting between the vector and 
pseudoscalar states. We showed in~\cite{Dowdall:2012ab} 
that improving $c_4$ leads to accurate results for $b$-light hyperfine splittings 
in keeping with the results of~\cite{Dowdall:2011wh} for bottomonium.
Most of the correlators we use here for determining the vector 
heavy-light meson decay constants come from the 
same calculation as that of~\cite{Dowdall:2012ab}. 

The tuning of the $b$ quark mass on these ensembles was 
discussed in~\cite{Dowdall:2011wh}. We use the spin-averaged kinetic 
mass of the $\Upsilon$ and $\eta_b$ and tune this to an experimental 
value of 9.445(2) GeV. 
This allows for electromagnetism 
and $\eta_b$ annihilation effects missing from our 
calculation~\cite{Gregory:2010gm}.
Note that we no longer have to apply a shift for missing charm quarks 
in the sea~\cite{Gregory:2010gm}. 
The values used in this calculation are the same as 
those in~\cite{Dowdall:2012ab, Dowdall:2013tga}
and given in table \ref{tab:upsparams} along with other parameters.

\begin{table}
\caption{ The coefficients $c_1$, $c_5$, $c_4$ and $c_6$ used 
in the NRQCD action (eq.~(\ref{eq:Hamiltonian})) for the 
values of the $b$ quark mass corresponding to those in~\ref{tab:upsparams}. 
$c_2$ and $c_3$ are set to 1.0.
}
\label{tab:wilsonparams}
\begin{ruledtabular}
\begin{tabular}{lllll}
Set & $c_1$ & $c_5$ & $c_4$ & $c_6$ \\
\hline
very coarse 	& 1.36 & 1.21 & 1.22 & 1.36 \\
coarse 		& 1.31 & 1.16 & 1.20 & 1.31 \\
fine 		& 1.21 & 1.12 & 1.16 & 1.21 \\
\end{tabular}
\end{ruledtabular}
\end{table}

\begin{table}
\caption[dbd]{ Parameters used in the NRQCD action.
$am_b$ is the 
bare $b$ quark mass and $u_{0L}$ the Landau link tadpole-improvement 
factor used in the NRQCD action \cite{Lepage:1992xa}.  
$\delta x_b$ gives the fractional mistuning in the $b$ quark mass
($(am_b-am_{b,\mathrm{phys}})/am_{b,\mathrm{phys}}$)
obtained from the determination of the spin-averaged kinetic 
mass of the $\Upsilon$ and $\eta_b$~\cite{Dowdall:2011wh}, when 
this has a magnitude larger than 0.5\%. 
The column $a_{\mathrm{sm}}$ gives the size parameters of the quark 
smearing functions (see section~\ref{subsec:2pt} and~\cite{Dowdall:2012ab}), 
which take the form $\exp(-r/a_{\mathrm{sm}})$. 
$a_{\mathrm{sm}}$ is kept approximately constant in physical units
{\footnote{Note that there was a typographical error in~\cite{Dowdall:2012ab}
in the table giving $a_{\mathrm{sm}}$ values for sets 5 and 6 - the correct values 
are the ones given here.}}. 
}
\label{tab:upsparams}
\begin{ruledtabular}
\begin{tabular}{lllll}
Set & $am_b$ & $\delta x_b$ & $u_{0L}$ & $a_{\mathrm{sm}}/a$  \\
\hline
1 & 3.297 & 0 & 0.81950  & 2.0,4.0  \\
2 & 3.263 & 0 & 0.82015  & 2.0,4.0 \\
3 & 3.25 & 0.005 & 0.81947 & 2.0,4.0  \\
\hline
4 & 2.66 & -0.013 & 0.8340   & 2.5,5.0 \\ 
5 & 2.62 & 0 & 0.8349  & 2.0,4.0 \\ 
6 & 2.62 & 0 & 0.8341  & 2.0,4.0 \\ 
\hline
7 & 1.91 & 0.009 & 0.8525 & 3.425,6.85  \\ 
\end{tabular}
\end{ruledtabular}
\end{table}

We end this section with a brief discussion of the assessment and removal 
of discretisation errors in a calculation that uses NRQCD~\cite{Dowdall:2011wh}. 
A typical procedure to remove finite-$a$ errors in a lattice QCD calculation 
consists of :
\begin{itemize}
\item assume that the error is given by a function with leading term $ca^2$ (for suitably 
accurate actions)
\item perform calculations at multiple values of $a$
\item determine the unknown parameter $c$ above by fitting the results as a function of $a$
\item subtract the fitted error function to obtain a physical result. 
\end{itemize}
The first step of the procedure changes for NRQCD, because the error function must be 
more complicated. The coefficient of $a^2$ errors will be in general a function $c(am_b)$. 
This function is not known but varies slowly with $am_b$ for $am_b > 1$. 
It can therefore be approximated by a simple polynomial in $am_b$ for the range 
of values of $am_b$ used here, which are all larger than 1. 
Note that this polynomial approximation is not valid as $am_b \rightarrow 0$, but 
the procedure only requires that it be valid over the range used for our results. 
Our fit to the $a$-dependence of our results, to be discussed further 
in Section~\ref{sec:results}, then has additional parameters 
to allow for the $a$-dependence coming from NRQCD 
(we also have simpler $a$-dependence
coming from the gluon and light quark actions). 
The final physical result then has larger uncertainties because of this 
but it does allow us to account for NRQCD effects.

%
\subsection{HISQ valence quarks}
\label{subsec:hisq}
%

For the $u/d$, $s$ and $c$ valence quarks in our calculation 
we use the same HISQ action as for the sea quarks. 
The advantage of using HISQ is that $am_q$ discretisation errors are under 
sufficient control that it can be used both for light 
and for $c$ quarks~\cite{Follana:2006rc, oldfds, newfds}. 
The HISQ action is also numerically inexpensive which 
means we are able to perform a very high statistics calculation 
to combat the signal to noise ratio problems that arise in simulating B-mesons. 
For example, we use 16 time sources for both NRQCD and HISQ propagators 
on each configuration, so we are typically generating 16,000 correlators per ensemble. 

The masses used on each ensemble are given in table~\ref{tab:hisqparams}. 
Again these are the same as in~\cite{Dowdall:2012ab, Dowdall:2013tga}. 
In~\cite{Dowdall:2011wh} accurate strange quark masses were 
determined for each ensemble, tuned from the mass of the $\eta_s$ meson, 
a pseudoscalar $s\overline{s}$ which can be prevented from mixing 
with other states on the lattice so that its mass can be determined 
very accurately~\cite{Davies:2009tsa}.  
Using experimental $K$ and $\pi$ meson masses we found 
$M_{\eta_s}$ = 0.6893(12) GeV (see also~\cite{Dowdall:2013rya}). 
The values of $am_s^{\mathrm{val}}$ in table \ref{tab:hisqparams} 
correspond to these tuned values. 
The light valence quarks are taken to have the same masses as in the sea. 

Charm quark masses are tuned by matching the mass of 
the $\eta_c$ to experiment. The experimental value is shifted 
by 2.6 MeV for missing electromagnetic effects and 2.4 MeV for not 
allowing it to annihilate to gluons, giving 2.985(3) GeV~\cite{Davies:2009tsa}. 
The $\epsilon_{\mbox{\tiny Naik}}$ term in the action is not 
negligible for charm quarks and we use the tree level formula 
given in \cite{newfds}; the values appropriate to our 
masses are given in table~\ref{tab:hisqparams}.

\begin{table}
\caption{ 
The parameters used in the generation of the HISQ propagators. 
$am_l^{\rm val}$ and $am_s^{\rm val}$ are the valence light and 
strange quark masses respectively, in lattice units. 
$am_c^{\rm val}$ is the charm quark mass in lattice units (only a subset 
of the ensembles was used in this case) and 
$\epsilon_{\mbox{\tiny Naik}}$ is the corresponding 
coefficient of the Naik term in the HISQ action for charm. 
On set 5 $\epsilon_{\mbox{\tiny Naik}}$ is very slightly 
wrong - it should be -0.224. The impact of this is negligible.   
\label{tab:hisqparams}
}
\begin{ruledtabular}
\begin{tabular}{lllll}
Set & $am_l^{\rm val}$ & $am_s^{\rm val}$ & $am_c^{\rm val}$ & $\epsilon_{\mbox{\tiny Naik}}$ \\
\hline
1 & 0.013   & 0.0641 & 0.826  & -0.345 \\
2 & 0.0064  & 0.0636 & 0.818  & -0.340\\
3 & 0.00235  & 0.0628 & -  & - \\
\hline
4 & 0.01044 & 0.0522 & 0.645  & -0.235  \\ 
5 & 0.00507 & 0.0505 & 0.627  & -0.222  \\ 
6 & 0.00184 & 0.0507 & -  & -  \\ 
\hline
7 & 0.0074  & 0.0364 & 0.434  & -0.117  \\ 
\end{tabular}
\end{ruledtabular}
\end{table}

%
\subsection{NRQCD-HISQ correlators}
\label{subsec:2pt}
%

The NRQCD $b$ and HISQ $u/d$, $s$ or $c$ light quark propagators 
are combined into a meson correlation function in a straightforward 
way. Since staggered quarks have no spin index, staggered quark 
propagators must first be converted to 4-component `naive' propagators 
so that they can be combined with quark propagators from other 
formalism such as NRQCD. To do this, the 4x4 `staggering matrix' 
$\Omega(x) = \prod_{\mu=0}^4 (\gamma_{\mu})^{x_{\mu}}$ 
that was used to convert the naive quark action into the 
staggered quark action has to be applied to the staggered quark propagator 
at each end to `undo' the transformation~\cite{Wingate:2002fh}. 
The spin and colour components of the 
naive propagator and the NRQCD propagator can then be tied up at source 
and sink with appropriate $\gamma$ matrices (taking appropriate 
$2\times 2$ blocks since the NRQCD propagator 
is 2-component) to form a 
pseudoscalar or vector meson correlator. 
We sum over the spatial sites on the sink time-slice to project onto zero 
spatial momentum in all cases. 

One complication is that `random-wall' sources 
(i.e. a set of U(1) random numbers over a timeslice) 
are used for the light quark propagators to improve statistical 
accuracy in our light meson calculations (see, for 
example,~\cite{Dowdall:2013rya}). 
When these propagators are tied together the result is 
equivalent to having a delta function source at each point 
on the time-slice. 
As well as the convenience, 
statistical accuracy is also improved by 
re-using these propagators in our heavy-light meson calculations. 
The source for the NRQCD 
propagators must then use the same random numbers on 
the same source time-slice and in addition must also 
include a spin trace over the staggering matrix and 
appropriate gamma matrix for a pseudoscalar or vector 
meson~\cite{Gregory:2010gm}, i.e. there is a separate NRQCD 
propagator for each meson that will be made.  
Combining these NRQCD propagators with the light quark propagators 
is then equivalent to having a delta function source at 
each point on the timeslice, as for the light meson case. 

A further numerical improvement is to make smeared sources 
for the NRQCD propagators by convoluting a smearing function with 
the random-wall source as above. Suitably chosen smearing functions 
can improve the overlap of the correlator with different states 
in the spectrum, and this is particularly important for 
fits to extract radially excited energies~\cite{Dowdall:2011wh}.
Here we use it to improve the determination of ground-state properties 
by improving the overlap with the ground-state at early times 
before the signal/noise ratio has degraded significantly. 
For each ensemble we then 
use a local source and 2 smeared sources. 
The smearing functions were optimised in our heavy-light meson 
spectrum calculation~\cite{Dowdall:2012ab} and take the 
form $\exp(-r/a_{sm})$ as a function of radial distance, 
with two different radial sizes, $a_{sm}$, on each 
ensemble as given in Table~\ref{tab:upsparams}.

Propagators were calculated, and meson correlators obtained, 
using 16 time sources on each configuration. The calculation was also 
repeated with the heavy quark propagating in the opposite time direction. 
All correlators from the same configuration were binned together 
to avoid underestimating the statistical errors.
When each smearing is used at source and sink this gives 
a $3\times 3$ matrix of correlation functions on each 
ensemble. In addition we have a 3-vector of correlation functions from 
using each smearing at the source and a relativistic current correction 
operator at the sink, to be discussed below.  

Meson energies and amplitudes 
are extracted from the meson correlation functions using 
a simultaneous multi-exponential Bayesian fit~\cite{gplbayes} as 
a function of time separation between source and sink to the form
\begin{eqnarray}
\label{eq:fitform}
C_{\rm meson}(i,j,t_0;t)&=& \sum^{N_{\exp}}_{k=0} b_{i,k}b^*_{j,k}e^{-E_{k}(t-t_0)
}\\
 &-& \sum^{N_{\exp}-1}_{k^\prime=0}   d_{i,k^\prime}d^*_{j,k^\prime}(-1)^{(t-t_0)
}e
^{-E^\prime_{k^\prime}(t-t_0)}.\nonumber
\end{eqnarray}
Here $i$, $j$ label the smearing (or current correction operator) included 
in the correlator and $k$ labels the set of energy levels 
for states appearing in the correlator. Here we are concentrating 
on the properties of the ground-state, $k=0$. 
$k^{\prime}$ labels a set of opposite-parity states that appear with an 
oscillating behaviour in time as a result of using staggered quarks. 
Energies for ground-states, radially excited states and oscillating states were 
extracted from these correlation functions in~\cite{Dowdall:2012ab}. 
Here we use similar fits to determine ground-state amplitudes and thereby decay 
constants.
 
The fits are straightforward and follow the same pattern in all 
cases. 
We fit the pseudoscalar and vector correlators simultaneously 
for each pair i.e. $B$ and $B^*$, $B_s$ and $B^*_s$ and 
$B_c$ and $B^*_c$. That enables us to extract a correlated 
ratio of amplitudes that we need for the ratio of decay constants. 
We take a prior on the ground-state energy determined 
from effective mass plots, 
with a width of 300 MeV. The prior on the lowest oscillating state 
is taken to be 400 MeV higher than the ground-state with a width 
of 300 MeV. The prior on the energy splittings in both the oscillating and 
non-oscillating sectors, 
$E_{n+1}-E_n$, is taken as 600(300) MeV and the priors on the 
amplitudes as 0.1(2.0). 
The fits include points from $t_{\mathrm{min}}$ 
to $t_{\mathrm{max}}$, close to half the temporal 
extent of the lattice. $t_{\mathrm{min}}$ 
is taken from $6-8$ for $B$ and $B_s$ fits
and 
$t_{\mathrm{max}}$ is taken 
as 18 on the very coarse lattices, 28 on coarse and 
40 on fine. For $B_c$ we use time ranges $12-24$ on very 
coarse, $8-21$ on coarse and $10-30$ on fine.   
In all cases we have good fits that reach stable ground-state 
parameters quickly. 
We take results from fits that 
use $N_{\mathrm{exp}}=4$. 
 
%
\subsection{Determining decay constants}
\label{subsec:decay}
%
Meson decay constants, $f$, are hadronic parameters 
defined from the matrix element of 
the local current that annihilates the meson (coupling for example 
to a $W$ boson).
For mesons at rest:
\begin{eqnarray}
\label{eq:fdef}
\langle 0 | J_{A_0} | H \rangle &=& f_H M_H \nonumber \\
\langle 0 | J_{V_i} | H_j^* \rangle &=& f_{H^*} M_{H^*} \delta_{ij} .
\end{eqnarray}
Here $H$ is one of the pseudoscalar mesons, $B_q$ for $q=l,s,c$. 
These matrix elements depend on the QCD interactions that keep 
the quark and antiquark bound inside the meson. They can be 
calculated directly from the amplitudes obtained from fits 
to the meson correlators, provided that we can accurately 
represent the continuum QCD currents, $J_{A_0}$ and $J_{V_i}$ 
on the lattice. 

The representation of these currents when combining a lattice 
NRQCD $b$-quark with a light quark is discussed most recently 
in~\cite{Dowdall:2013tga, Monahan:2012dq}. 
The procedure is similar for the temporal axial and spatial vector 
currents and so we just give the temporal axial case in detail. 

For the temporal axial current whose matrix element gives the 
pseudoscalar decay constant, we determine matrix elements on 
the lattice made from light quark fields $\Psi_q$ and NRQCD 
field $\Psi_Q$ of:
\begin{eqnarray}
\label{eq:jpsdef}
J_{A_0}^{(0)} &=& \overline{\Psi}_q \gamma_5 \gamma_0 \Psi_Q \\
J_{A_0}^{(1)} &=& -\frac{1}{2m_b}\overline{\Psi}_q \gamma_5 \gamma_0 \vec{\gamma}\cdot\vec{\nabla} \Psi_Q .\nonumber 
\end{eqnarray} 
$J_{A_0}^{(0)}$ is the leading term in a nonrelativistic expansion of 
the current operator, and $J_{A_0}^{(1)}$ is the first relativistic 
correction, appearing with one inverse power of the $b$ quark mass. 

$J_{A_0}^{(0)}$ is simply the operator that corresponds to our local 
sources for the $b$ quark described in Section~\ref{subsec:2pt}. 
Thus the matrix element of $J_{A_0}^{(0)}$ 
between the vacuum and the ground-state meson 
is obtained directly from the amplitude of this operator from 
our fit function, i.e. $b_{\mathrm{loc},0}$ from eq.~(\ref{eq:fitform}). 
By inserting a complete set of states with standard normalisation 
into the pseudoscalar meson correlation function we have 
\begin{equation}
\label{eq:amptophi}
b_{\mathrm{loc},0} = \frac{\langle 0 | J_{A_0}^{(0)} | H \rangle } {\sqrt{2M_H}} .
\end{equation}
Similarly, by inserting the operator $J_{A_0}^{(1)}$ at the sink 
for meson correlators made from the three different sources that we use 
we can determine an 
amplitude for this operator in the ground-state
\begin{equation}
\label{eq:ampj1}
b_{J1,0} = \frac{\langle 0 | J_{A_0}^{(1)} | H \rangle } {\sqrt{2M_H}} .
\end{equation}

The way in which $J_{A_0}^{(0)}$ and $J_{A_0}^{(1)}$ can be 
combined into an 
accurate representation of $J_{A_0}$ from full QCD is 
described in~\cite{Dowdall:2013tga}. Here, for most of our results, 
we will use an expression that is slightly less accurate 
than that in~\cite{Dowdall:2013tga}. We take:   
\begin{equation}
\label{eq:jcomb}
J_{A_0} = (1+z_{A_0}\alpha_s)(J_{A_0}^{(0)} + J_{A_0}^{(1)}) .
\end{equation}
Thus, we can combine matrix elements above to obtain: 
\begin{eqnarray}
\label{eq:phidef}
\Phi_{A_0}^{(0)} &=& \sqrt{2} b_{\mathrm{loc},0} \\ 
\Phi_{A_0}^{(1)} &=& \sqrt{2} b_{J1,0} \nonumber \\ 
f_H\sqrt{M_H} &=&  (1+z_{A_0}\alpha_s)(\Phi_{A_0}^{(0)} + \Phi_{A_0}^{(1)}) \nonumber
\end{eqnarray}
up to sources of uncertainty that will be discussed in the appropriate 
subsections of Section~\ref{sec:results}. 
Note that the decay constant appears 
naturally multiplied by the square root of the meson mass in these 
expressions.  

Analogous expressions are used for the vector current case, using amplitudes
from the vector meson correlator fits. 

Systematic errors are reduced by working with 
the ratio of vector to pseudoscalar meson 
decay constants (multiplied by the ratio of the square root of 
the masses). Hence we define the quantity $R_q$ for meson $B_q$, determined from:  
\begin{equation}
\label{eq:frat}
R_q \equiv \frac{f_H^*\sqrt{M_H^*}}{f_H\sqrt{M_H}} = (1 + \delta z \cdot \alpha_s) \frac{(\Phi_{V_i}^{(0)} + \Phi_{V_i}^{(1)})}{(\Phi_{A_0}^{(0)}+\Phi_{A_0}^{(1)})} .
\end{equation}
For convenience we expand the ratio of renormalisation constants 
to $\mathcal{O}(\alpha_s)$ so that 
$\delta z$ is $z_{V_i}-z_{A_0}$. $\delta z$ will be tabulated, along with 
the results, in Section~\ref{sec:results}. This expression is accurate 
up to missing $\alpha_s^2$ pieces of the overall renormalisation factor 
(i.e. in the term $(1+\delta z \cdot \alpha_s)$)
and missing additional $\alpha_s$ renormalisation factors for the 
sub-leading current contributions (that would appear multiplying 
$\Phi_{A_0}^{(1)}$ for example). These sources of systematic uncertainty 
will be estimated in Section~\ref{sec:results} and included in our 
final error budgets. 

We will first calculate $R_s$ as the `calibration' ratio of vector 
to pseudoscalar decay constants. It is convenient subsequently to 
calculate ratios of $R_l$ and $R_c$ to $R_s$. Some systematic errors 
cancel in these ratios of ratios, allowing us to obtain a more accurate 
picture of how much the ratio of vector to pseudoscalar heavy-light 
meson decay constants depends on the light quark mass. 

\section{Results}
\label{sec:results}
\subsection{$b$-light correlators}
\label{subsec:BBs}
\begin{table}
\caption{Amplitudes for $J^{(0)}$ and $J^{(1)}$ for 
temporal axial and vector currents between the vacuum 
and the $B_s$ and $B^*_s$ mesons respectively, extracted 
from correlator fits and multiplied by $\sqrt{2}$ in accordance 
with eq.~(\ref{eq:phidef}). 
Results are in lattice units and the errors given are statistical/fit 
errors only. Results for the $B_s$ were previously given 
in~\cite{Dowdall:2013tga}. Results here differ slightly because the 
fits included both vector and pseudoscalar correlators in a simultaneous 
fit and also incorporated more correlators that included $J^{(1)}$ amplitudes. 
\label{tab:phibs}
}
\begin{ruledtabular}
\begin{tabular}{lllll}
Set & $a^{3/2}\Phi_{B_s}^{(0)}$ & $a^{3/2}\Phi_{B_s}^{(1)}$ & $a^{3/2}\Phi_{B^*_s}^{(0)}$ & $a^{3/2}\Phi_{B^*_s}^{(1)}$ \\
\hline
1 & 0.3714(8)   & -0.02939(10) & 0.3403(12) & 0.00909(4) \\
2 &  0.3628(13) & -0.02874(13) & 0.3321(10) & 0.00889(3)\\
3 & 0.3606(9)  & -0.02870(9) & 0.3295(4) & 0.00887(1) \\
\hline
4 & 0.2728(5) & -0.02343(6) & 0.2425(7) & 0.00706(3)  \\ 
5 & 0.2680(3) & -0.02323(4) & 0.2369(5) & 0.00697(2)  \\ 
6 & 0.2657(2) & -0.02298(2) & 0.2351(2) & 0.00689(1) \\ 
\hline
7 & 0.1747(2) & -0.01713(3) & 0.1491(3) & 0.00497(1)  \\ 
\end{tabular}
\end{ruledtabular}
\end{table}

\begin{table}
\caption{Amplitudes for $J^{(0)}$ and $J^{(1)}$ for 
temporal axial and vector currents between the vacuum 
and the $B_l$ and $B_l^*$ mesons respectively, extracted 
from correlator fits and multiplied by $\sqrt{2}$ in accordance 
with eq.~(\ref{eq:phidef}). $l$ denotes a $u$ or $d$ quark, 
taken here to have the same mass.  
Results are in lattice units and the errors given are statistical/fit 
errors only. Results for the $B_l$ were previously given 
in~\cite{Dowdall:2013tga}. Results here differ slightly for reasons given
in the caption to Table~\ref{tab:phibs}. 
\label{tab:phib}
}
\begin{ruledtabular}
\begin{tabular}{lllll}
Set & $a^{3/2}\Phi_{B_l}^{(0)}$ & $a^{3/2}\Phi_{B_l}^{(1)}$ & $a^{3/2}\Phi_{B^*_l}^{(0)}$ & $a^{3/2}\Phi_{B^*_l}^{(1)}$ \\
\hline
1 &  0.3245(20)  & -0.02612(21) & 0.2964(24) & 0.00812(9) \\
2 & 0.3062(21)  & -0.02456(25) & 0.2752(29) & 0.00748(9)\\
3 & 0.2962(37)  & -0.02381(30) & 0.2681(31) &  0.00719(13)\\
\hline
4 & 0.2352(21) & -0.02033(21) & 0.2086(25) &  0.00623(12) \\ 
5 & 0.2276(13) & -0.01989(15) & 0.1997(16) &  0.00596(6) \\ 
6 & 0.2190(14) & -0.01904(16) & 0.1915(20) & 0.00558(9) \\ 
\hline
7 & 0.1521(4) & -0.01500(5) & 0.1292(4) & 0.00432(2)  \\ 
\end{tabular}
\end{ruledtabular}
\end{table}

\begin{table}
\caption{Coefficients $z_{A_0}$ and $z_{V_i}$ 
needed for the one-loop renormalisation factor 
for the pseudoscalar and vector decay constants 
respectively for the values of 
$m_ba$ used on the different ensembles. 
$z$ is constructed from results 
given for the appropriate NRQCD bare masses and 
massless HISQ quarks in~\cite{Monahan:2012dq} 
as $z = \rho_0-\zeta_{10}$. The uncertainties come 
from statistical errors in the numerical integration, 
taken to be uncorrelated. 
In~\cite{Dowdall:2013tga} 
$z_{A_0}$ is called $z_0$.  
Column 4 gives $\delta z$ which is the difference 
between $z_{V_i}$ and $z_{A_0}$.  
Column 5 gives the corresponding values of 
$\delta z$ for the case where only the leading-order 
NRQCD currents ($J^{(0)}_{A_0}$ and $J^{(0)}_{V_i}$) 
are used in the calculation. 
\label{tab:zlight}
}
\begin{ruledtabular}
\begin{tabular}{lllll}
$m_ba$ & $z_{V_i}$ & $z_{A_0}$ & $\delta z$ & $\delta z^{\mathrm{LO}}$ \\
\hline
3.297 &  -0.078(2)  & 0.024(2) & -0.102(3) &  0.026(3) \\
3.263 & -0.077(2)  & 0.022(2) &  -0.099(3) & 0.030(3) \\
3.25 & -0.077(2)  & 0.022(1) &  -0.099(3) &  0.030(3) \\
\hline
2.66 & -0.073(2) & 0.006(2) &  -0.079(3) &  0.076(3) \\ 
2.62 & -0.072(2) & 0.001(2) &  -0.073(3) &  0.083(3) \\ 
\hline
1.91 & -0.044(2) & -0.007(2) &   -0.037(3) &  0.168(3)\\ 
\end{tabular}
\end{ruledtabular}
\end{table}

\begin{table}
\caption[kldgrd]{Results for ratios of amplitudes 
for vector and pseudoscalar mesons on each ensemble as defined in the 
text. 
Column 2 gives the unrenormalised ratio of amplitudes for the $B_s^*/B_s$ including 
the current corrections, 
$R_s^{\mathrm{unren.}} = (\Phi^{(0)}_{B^*_s}+\Phi^{(1)}_{B^*_s})/(\Phi^{(0)}_{B_s}+\Phi^{(1)}_{B_s})$ 
eq.~(\ref{eq:unrenrat}). 
Column 3 gives the equivalent quantity 
for the $B^*_l/B_l$ mesons. 
Column 4, $R_s^{\mathrm{LO}}$, gives 
renormalised ratio from eq~(\ref{eq:fratlo}) but including only 
the leading-order NRQCD currents, $J^{(0)}$.  
Finally column 5 gives the renormalised ratio of amplitudes including 
the current corrections. These numbers are determined 
from eq.~(\ref{eq:frat}) and plotted as the points in Figure~\ref{fig:Bsrat}. 
$R_s = (f_{B^*_s}\sqrt{M_{B^*_s}})/(f_{B_s}\sqrt{M_{B_s}})$, 
correct through $\mathcal{O}(\alpha_s)$ and 
$\mathcal{O}(\Lambda/m_b)$. 
Errors on the values are statistical only, but include correlations 
between vector and pseudoscalar meson correlation functions.   
\label{tab:compbs}
}
\begin{ruledtabular}
\begin{tabular}{lllll}
Set & $R_s^{\mathrm{unren.}}$ & $R_l^{\mathrm{unren.}}$ & $R_s^{\mathrm{LO}}$ & $R_s$ \\
\hline
1 & 1.0215(25) &  1.0205(59) & 0.9243(24)  & 0.9854(26)  \\
2 & 1.0206(36)  &  1.0038(73) & 0.9247(34) & 0.9858(36)\\
3 & 1.0196(26)  &  1.0106(97) & 0.9232(25) & 0.9850(27) \\
\hline
4 & 1.0006(19) & 1.0001(90) & 0.9098(19) & 0.9760(21) \\ 
5 & 0.9965(17) & 0.9899(71) & 0.9067(18) &  0.9741(20) \\ 
6 & 0.9967(8) & 0.9854(97) & 0.9071(11) & 0.9744(12) \\ 
\hline
7 & 0.9775(14) & 0.9734(23)  & 0.8916(16) & 0.9678(17) \\ 
\end{tabular}
\end{ruledtabular}
\end{table}

Correlators for $B_s$, $B^*_s$, $B_l$ and $B^*_l$ are fitted 
as described in Section~\ref{sec:lattice} and results for 
the ground-state amplitudes of leading, $J^{(0)}$, 
and sub-leading, $J^{(1)}$, currents 
are tabulated in Tables~\ref{tab:phibs} and~\ref{tab:phib} 
respectively. In Table~\ref{tab:compbs} we also tabulate for 
both $B_s$ and $B_l$ the 
ratio of the sum of the amplitudes that make up the NRQCD vector 
and temporal axial currents (without any renormalisation factors) 
defined as:  
\begin{equation}
\label{eq:unrenrat}
R_q^{\mathrm{unren.}} \equiv \frac{(\Phi_{V_i}^{(0)} + \Phi_{V_i}^{(1)})}{(\Phi_{A_0}^{(0)}+\Phi_{A_0}^{(1)})} .
\end{equation}
These ratios are determined directly from the fits, including the 
correlations between the fitted amplitudes for vector and 
pseudoscalar mesons, 
and therefore have 
smaller statistical errors than determining them naively from the 
results in Tables~\ref{tab:phibs} and~\ref{tab:phib}.  

The $z$ factors needed to multiply $\alpha_s$ in the one-loop 
renormalisation for the temporal axial and spatial 
vector currents are given in Table~\ref{tab:zlight}. These 
are calculated for massless HISQ light quarks and the 
values of $am_b$ in the NRQCD action used on each of 
the ensembles. The fact that these $z$ coefficients are very 
small was already noted in~\cite{Dowdall:2013tga}. This 
means that renormalisation factors to the continuum 
current are close to 1~\footnote{Note that we do not need an 
initial nonperturbative step to achieve this, as is 
used by the Fermilab Lattice/MILC Collaboration~\cite{Harada:2001fi}. 
That step is largely required to remove large but 
generic renormalisation factors associated with the clover 
action and has been tested nonperturbatively in~\cite{Chakraborty:2014zma}. }.

\begin{figure}[t]
\includegraphics[width=0.9\hsize]{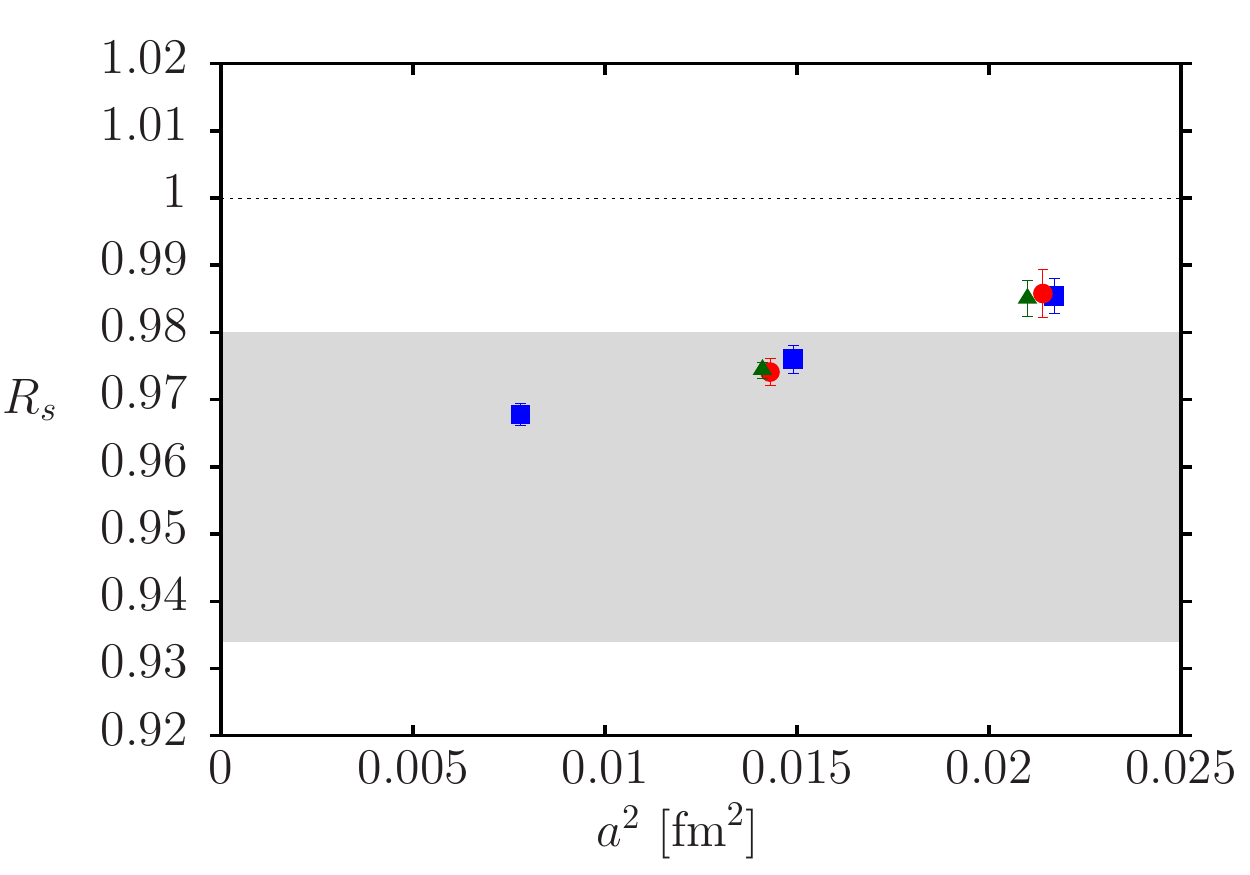}
\caption{Results for $R_s$, the ratio of $B^*_s$ to $B_s$ decay constants 
(multiplied by the square root of the mass ratio), plotted against 
the square of the lattice spacing in $\mathrm{fm}^2$. 
Note the magnified $y$-axis scale. 
The errors on the data points include statistical/fitting errors. 
Blue filled squares are results on sets with $m_l/m_s=0.2$, red 
filled circles sets with $m_l/m_s=0.1$ and green filled triangles 
sets with physical $m_l$. 
The grey shaded band gives our physical result 
including all systematic errors discussed in the text.
The black dotted line marks the value 1.0. 
}
\label{fig:Bsrat}
\end{figure}

From Table~\ref{tab:phibs} and~\ref{tab:phib} it is 
immediately clear that the leading order amplitudes, $\Phi^{(0)}$, 
show a difference between vector and pseudoscalar mesons 
with the vector result being smaller than the pseudoscalar. 
This difference is largely down to the `hyperfine' 
interaction in the NRQCD Hamiltonian (the term with 
coefficient $c_4$ in eq.~(\ref{eq:Hamiltonian})).  
The Tables make clear that the impact of this interaction 
is to lower the ratio of vector to pseudoscalar decay constants. 
This effect agrees in sign with that seen in an earlier 
lattice NRQCD analysis of the impact of different relativistic 
corrections on heavy-light meson decay constants~\cite{sara, Collins:1999ff}. 
It also agrees with early estimates using HQET and QCD sum rules~\cite{Ball:1994uh}. 

From the tables it is also clear that the relativistic current correction 
matrix element, $\Phi^{(1)}$, has opposite sign for the 
vector and pseudoscalar cases, being positive for the vector 
and negative for the pseudoscalar. The impact of these corrections 
is then to raise the ratio of vector to pseudoscalar decay 
constants. This sign, and the fact that the pseudoscalar $J^{(1)}$ matrix 
element is approximately three times that of the vector, 
agrees with HQET expectations~\cite{Neubert:1992fk, Ball:1994uh} and 
earlier lattice NRQCD analyses~\cite{sara, Collins:1999ff}. 

Simply dividing the current correction matrix element, $\Phi^{(1)}$, 
by $\Phi^{(0)}$ gives naively a relative contribution to the amplitude from the 
relativistic current corrections of size -(8-10)\% for the pseudoscalar and
+3\% for the vector. 
This does not take into account the fact 
that the addition of the relativistic current correction $J^{(1)}$ 
which appears at tree-level 
changes the overall renormalisation of the lattice NRQCD current at 
$\mathcal{O}(\alpha_s)$ because radiative corrections 
to $J^{(1)}$ can look like $J^{(0)}$~\cite{Monahan:2012dq}. 
Therefore to determine 
more accurately the effect of the relativistic current corrections 
we have to compare the renormalised result with and without the 
inclusion of the $J^{(1)}$ current correction. 

This is done in Table~\ref{tab:compbs} 
in which we compare the two results for the 
ratio of $f\sqrt{M}$ for $B^*_s$ and $B_s$. 
The right-hand column, denoted $R_s$, is the full 
result obtained from eq.~(\ref{eq:frat}) using the 
values of $\delta z$ from Table~\ref{tab:zlight}. 
The values 
for $\alpha_s$ used in that expression are taken in 
$V$ scheme at the scale $2/a$, where $a$ is the lattice 
spacing on that ensemble, and are
given in Table~\ref{tab:params}. 
The 
column denoted $R_s^{\mathrm LO}$ gives results using 
only the leading-order currents and 
\begin{equation}
\label{eq:fratlo}
R^{\mathrm{LO}}_s = (1 + \delta z^{\mathrm{LO}} \cdot \alpha_s) \frac{\Phi_{B^*_s}^{(0)} }{\Phi_{B_s}^{(0)}} 
\end{equation}
with $\delta z^{\mathrm{LO}}$ values given in Table~\ref{tab:zlight} and 
the same values of $\alpha_s$. Note 
the difference between $\delta z^{\mathrm{LO}}$ and $\delta z$. 
Both coefficients are small, but they have opposite sign. 
This then compensates to some extent 
for the effect of the current corrections and 
means that, comparing $R_s$ and $R_s^{\mathrm{LO}}$ 
in Table~\ref{tab:compbs} we 
see now that the total effect of the current correction terms in the 
ratio amounts to 7-8\%, 
somewhat less than the naive estimate of 12-15\%. There is of course an uncertainty 
on this estimate coming from missing $\alpha_s^2$ terms in the 
renormalisation. A similar procedure would be needed to estimate accurately the 
effect of the hyperfine term on the ratio $R_q$. However, because the 
hyperfine interaction is embedded in the NRQCD Hamiltonian it is automatically 
included in the perturbative matching calculation for the NRQCD currents 
and we do not have the $z$ coefficients without the hyperfine term included.  
Note that the size of the hyperfine coefficient 
($c_4$ in eq.~(\ref{eq:Hamiltonian})) is tested through determination 
of the mass splitting between vector and pseudoscalar mesons in~\cite{Dowdall:2012ab}.

\subsection{$f_{B_s^*}$}
\label{subsec:bsstar}
Figure~\ref{fig:Bsrat} plots 
the full results for $R_s$, the ratio of 
$f\sqrt{M}$ for the $B^*_s$ and $B_s$ 
mesons, obtained from eq.~(\ref{eq:frat}) and given as 
column 5 of Table~\ref{tab:compbs}. 
Statistical 
errors in $R_s$ are small, less than 0.5\%, so we see that the value 
for the ratio is clearly less than 1 and the dependence on 
the lattice spacing is small, but clear and unambiguous.  
To derive a physical result we need to fit this dependence, 
as discussed in Section~\ref{subsec:nrqcd},
allowing for other systematic uncertainties from lattice QCD. 

The key sources of systematic error that need 
to be allowed for, by inclusion in our fit function, are: 
\begin{itemize}
\item {\it Matching uncertainties - $\alpha_s^2$}. The missing $\alpha_s^2$ 
coefficient in the overall renormalisation factor for the 
ratio of amplitudes of the NRQCD currents is potentially the 
largest source of uncertainty here. We can allow for this by 
simply taking a fractional error which is $\alpha_s^2 \approx 0.1$ times 
a value for this coefficient. However, the value of 
$\alpha_s^2$ changes with the lattice spacing and the 
coefficient may also depend on $am_b$, as the known one-loop 
coefficient $\delta z$ does, see Table~\ref{tab:zlight}. 
Thus a better estimate is obtained by incorporating a factor 
to take account of this missing term into the fit. 
We write the factor as $(1+c\alpha_s^2)$ and take $c$ to 
have the form $c_1\times(1+c_2 \delta x_m + c_3 \delta x_m^2)$ 
where $c_1$ sets the overall allowed size of the coefficient 
and the $\delta x_m$ terms allow for dependence on $am_b$. 
$\delta x_m = (am_b-2.7)/1.5$ varies from -0.5 to 0.5 over the 
range of $am_b$ values we use here~\cite{Dowdall:2011wh}. 
\item {\it Matching uncertainties - $\alpha_s\Lambda/m_b$}. 
We must also allow for missing $\alpha_s$ terms that alter the 
normalisation of the relativistic current corrections 
within the NRQCD current and/or include the matrix elements of 
additional current corrections that only appear first at 
$\mathcal{O}(\alpha_s\Lambda/m_b)$. Such corrections 
were included in our determination of $f_{B_s}$ 
and $f_B$ in~\cite{Dowdall:2013tga} since they are known for the 
temporal axial current for massless HISQ quarks~\cite{Monahan:2012dq}.  
They will be discussed further in Subsection~\ref{subsec:mQ} but 
here we must include an uncertainty for the fact that they 
are missing in our ratio. For this we can include an additional 
term in the factor described above of the form $d\alpha_s\Lambda/m_b$ 
where $d$ has an expansion in powers of $\delta x_m$ of the 
same form as $c$ above. Here we can take $\Lambda/m_b$ to be 
0.08, the size of the relativistic current corrections as determined above. 
\item{\it Matching uncertainties - $(\Lambda/m_b)^2$}. Further 
current corrections at the next order in the relativistic 
expansion would appear at $(\Lambda/m_b)^2$. Since we have no 
information about these we do not include them in the fit  
but take an additional uncertainty of $(0.1)^2$ = 1\% (where 
0.1 is a suitable power-counting estimate of $\Lambda/m_b$) 
to account for them.  
\item{\it NRQCD systematics}. The improved NRQCD Hamiltonian 
that we use (eq.~(\ref{eq:Hamiltonian})) is accurate 
through $\mathcal{O}(\alpha_s\Lambda/m_b)$ 
in the context of heavy-light power-counting. Thus the hyperfine 
interaction that contributes to $R_s$ is accurate through this 
order, which is to a higher order than the matching uncertainties 
discussed above. 
Errors from the NRQCD 
Hamiltonian are then smaller than, and are effectively included in, 
the matching uncertainties already discussed. 
Likewise missing terms in the 
NRQCD Hamiltonian are at even higher order, 
$\mathcal{O}(1/m_b^3)$~\cite{Lepage:1992tx}.  
\item{\it Discretisation uncertainties}. These can come from 
the gluon action, the HISQ action and the NRQCD action. 
However, most discretisation uncertainties will cancel 
between vector and pseudoscalar mesons since the difference 
between them is a spin-dependent effect and hence suppressed 
by $\Lambda/m_b$. This is clear from Figure~\ref{fig:Bsrat} 
which shows very little dependence on $a$.  
In all three actions discretisation errors appear as 
even powers of $a$. We therefore include a factor 
$(1+(\Lambda/m_b)\sum_j e_j (\Lambda a)^{2j})$ to allow 
for these uncertainties in the fit. We take a value 0.2 
for $\Lambda/m_b$ here to be conservative. For $e_1$ we allow for dependence 
on $am_b$ coming from the NRQCD action as discussed in Section~\ref{subsec:nrqcd} 
and above 
for the coefficients $c$ and $d$.   
\item{\it Tuning uncertainties - valence quark masses}. Our valence 
masses are tuned very accurately (to an uncertainty of 1\%) but we 
allow for effects of mistuning. For the $s$ quark mass these 
will be negligible since, as we show below, the difference 
between $R_s$ and $R_l$ is very small. Mistuning of the 
$b$ quark mass will affect $R_s$ through the hyperfine 
interaction and the size of the current correction matrix 
elements, i.e. through a term of the form $(\Lambda/m_b) \delta m_b/m_b$. 
We therefore allow for a term of this form 
in the factor that includes discretisation 
effects above. We determine $\delta m_b$ from the physical 
values for $m_b$ given on each ensemble 
in~\cite{Dowdall:2011wh, Dowdall:2012ab} and these 
are tabulated in Table~\ref{tab:upsparams}. The largest 
value of $\delta m_b/m_b$ is 1.3\% on set 4. 
\item{\it Tuning uncertainties - sea quark masses}. Our results 
include values on ensembles of gluon field configurations 
at a variety of values of the $u/d$ quark mass in the sea, 
varying from $0.2m_s$ down to the physical point. 
The $s$ and $c$ quark masses in the sea are well-tuned. 
Dependence on the sea quark masses is very small, as is clear 
from Figure~\ref{fig:Bsrat}. We therefore include a simple linear 
dependence on the sea quark masses, as might be expected 
from leading-order chiral perturbation theory. This 
dependence takes the form 
$g \delta m_{\mathrm{sea}}/(10 m_{\mathrm{sea, phys}})$ 
where the mass-dependent variable is a physical one because 
we take a mass ratio in which $Z$ factos cancel. 
We include $u/d$ and $s$ quarks in $m_{\mathrm{sea}}$ and the 
factor of 10 is a convenient way to introduce the chiral scale 
of 1 GeV expected from chiral perturbation theory. 
$\delta m_{\mathrm{sea}} = (2m_l+m_s)-(2m_{l,\mathrm{phys}}+m_{s,\mathrm{phys}})$
and is obtained using values for $m_{s,\mathrm{phys}}$ given in~\cite{Dowdall:2011wh}. 
We take $m_{l,\mathrm{phys}}/m_{s,\mathrm{phys}}$ = 1/27.4~\cite{Bazavov:2014wgs}. 
Values for $\delta x_{\mathrm{sea}} \equiv \delta m_{\mathrm{sea}}/m_{\mathrm{sea, phys}}$
on each ensemble are given in Table~\ref{tab:params}.
\item{\it Uncertainties in the value of the lattice spacing}. Since we are determining 
a dimensionless ratio of decay constants, uncertainties in the 
value of the lattice spacing only enter indirectly through the uncertainty 
in tuning the quark masses. As discussed above the tuning of $m_b$ affects the 
size of the relativistic correction terms that affect the 
vector/pseudoscalar ratio. 
We have a 1\% uncertainty in our lattice spacing 
values, largely correlated between the ensembles and so we add 
an additional overall uncertainty of $0.2\times0.01$ = 0.2\% to allow 
for this. The factor of 0.2 is a conservative estimate for the 
size of relativistic corrections.  
\end{itemize}

Putting the features above together we arrive at a fit 
form for $R_s$ as a function of $a$ and quark masses as: 
\begin{eqnarray}
\label{eq:fitrs}
R_s(a,m) &=& R_{s,\mathrm{phys}}\times F_1(a,m) /F_2(\alpha_s) \\
F_2(\alpha_s) &=& (1 + c\alpha_s^2 + 0.08d\alpha_s) \nonumber \\
F_1(a,m) &=& 1 + 0.2\sum_{j=1}^3 e_j (\Lambda a)^{2j} \nonumber \\  
&+& 0.2 f \frac{\delta m_b}{m_{b,\mathrm{phys}}} + g \frac{\delta m_{\mathrm sea}}{10m_{\mathrm{sea,phys}}}  \nonumber \\
c &=& c_1\times(1+c_2\delta x_m + c_3 (\delta x_m)^2) \nonumber \\
d &=& d_1\times(1+d_2\delta x_m + d_3 (\delta x_m)^2) \nonumber \\
e_1 &=& e_{11}\times(1+e_{12}\delta x_m + e_{13} (\delta x_m)^2) \nonumber 
\end{eqnarray}
In dividing by $F_2$ we follow the convention that we used 
at $\mathcal{O}(\alpha_s)$ in eq.~(\ref{eq:frat}) of writing 
the renormalisation as a multiplicative factor.  
Thus if $F_2$ were instead known, rather than fitted, the raw 
results would be multiplied by this correction factor along 
with the factor at $\mathcal{O}(\alpha_s)$. 
We use a Bayesian fitting approach~\cite{gplbayes} 
to implement the fit function of eq.~(\ref{eq:fitrs}). Priors 
on all of the coefficients are taken as 0.0(1.0) except for $c_1$, 
which is taken as 0.0(0.2). This allows for an $\alpha_s^2$ coefficient 
in the overall renormalisation factor that is twice as large as 
the largest seen at $\mathcal{O}(\alpha_s)$ (see $\delta z$ values 
in Table~\ref{tab:zlight}). The prior on the physical value, $R_{s,\mathrm{phys}}$, 
is taken as 1.0(0.2). 

Applying this fit function to our results gives a $\chi^2/{\mathrm{dof}}=0.13$ 
and a physical result for $R_s$ of 0.957(23), when we include the uncertainty 
from missing higher order current corrections and the lattice spacing. 
The error budget from the fit is 
laid out in Table~\ref{tab:err}. As expected the uncertainty is dominated by 
that from current matching, although the fit has constrained this 
uncertainty to be a bit smaller than the naive expectation. 
The physical value, along with the total error, 
is plotted as a grey band on Figure~\ref{fig:Bsrat}.  
$R_s$ is the ratio of decay constants multiplied by the square root 
of the meson masses. Our earlier results~\cite{Dowdall:2012ab} 
showed that the vector and pseudoscalar meson masses 
calculated here agree with experiment. 
We can therefore convert our value of $R_s$ to 
a ratio for the decay constants 
using the square root of the experimental ratio of the 
meson masses of 1.0045(2)~\cite{pdg}. We obtain:
\begin{equation}
\label{eq:resrs}
\frac{f_{B^*_s}}{f_{B_s}} = 0.953(23) .
\end{equation} 
This is 2$\sigma$ below 1. 

\begin{table}[ht]
\caption{Full error budget for the various ratios of vector to pseudoscalar 
decay constants that we calculate here, giving each error 
as a percentage of the final answer, following the discussion of 
uncertainties in the text. The effects of finite volume and missing electromagnetism 
are expected to be negligible. 
   }
\label{tab:err}
\begin{ruledtabular}
\begin{tabular}{llll}
& $R_s$ & $R_l/R_s$ & $R_c/R_s$ \\
\hline
stats/fitting/scale	& 0.6	& 0.8 	&  0.7   \\
current matching        &  1.9     &  1.0    &  0.8  \\
$(\Lambda/m_b)^2$ currents   &  1.0     & 0.2    & 0.5 \\
$a$-dependence 		& 0.9 	& 0.1	&  0.15   \\
$m_{\mathrm{sea}}$-dependence 	& 0.05	& 0.2 	&  0.1   \\
$m_b$ tuning	 	& 0.4	& 0.03      &  0.1 \\
\hline
Total 			& 2.4 	& 1.3	&   1.2 
\end{tabular}
\end{ruledtabular}
\end{table}

\subsection{$f_{B^*}$}
\label{subsec:bstar}
To analyse the corresponding ratio, $R_l$, for the $B/B^*$ 
it is convenient to take the ratio to $R_s$. 
Table~\ref{tab:phib} gives our results for the $B$ and 
$B^*$ amplitudes and Table~\ref{tab:compbs} gives the 
ratio of the sum of amplitudes for $J^{(0)}$ and $J^{(1)}$ 
for vector and pseudoscalar. These results include the 
correlations between the vector and pseudoscalar meson correlators
from the simultaneous fit. 
The results for $B^*_l/B_l$ are very similar, not surprisingly, to 
those for $B_s$ and $B^*_s$. The statistical errors 
are significantly larger, however, as is expected 
when the light quark mass is reduced~\cite{newfds}. 
The renormalisation factor (eq.~(\ref{eq:frat})) for $R_l$ is the same as 
that for $R_s$ (since mass effects for 
light quarks are negligible in the matching) 
and so the renormalisation cancels in 
the ratio $R_l/R_s$. We can therefore simply determine 
$R_l/R_s$ from the ratio of the first two columns 
in Table~\ref{tab:compbs}: 
\begin{equation}
\label{eq:rlrsdef}
\frac{R_l}{R_s} = \left(\frac{\Phi^{(0)}_{B^*_l}+\Phi^{(1)}_{B^*_l}}{\Phi^{(0)}_{B_l}+\Phi^{(1)}_{B_l}}\right)\left(\frac{\Phi^{(0)}_{B_s}+\Phi^{(1)}_{B_s}}{\Phi^{(0)}_{B^*_s}+\Phi^{(1)}_{B^*_s}}\right) .
\end{equation}

\begin{figure}[t]
\includegraphics[width=0.9\hsize]{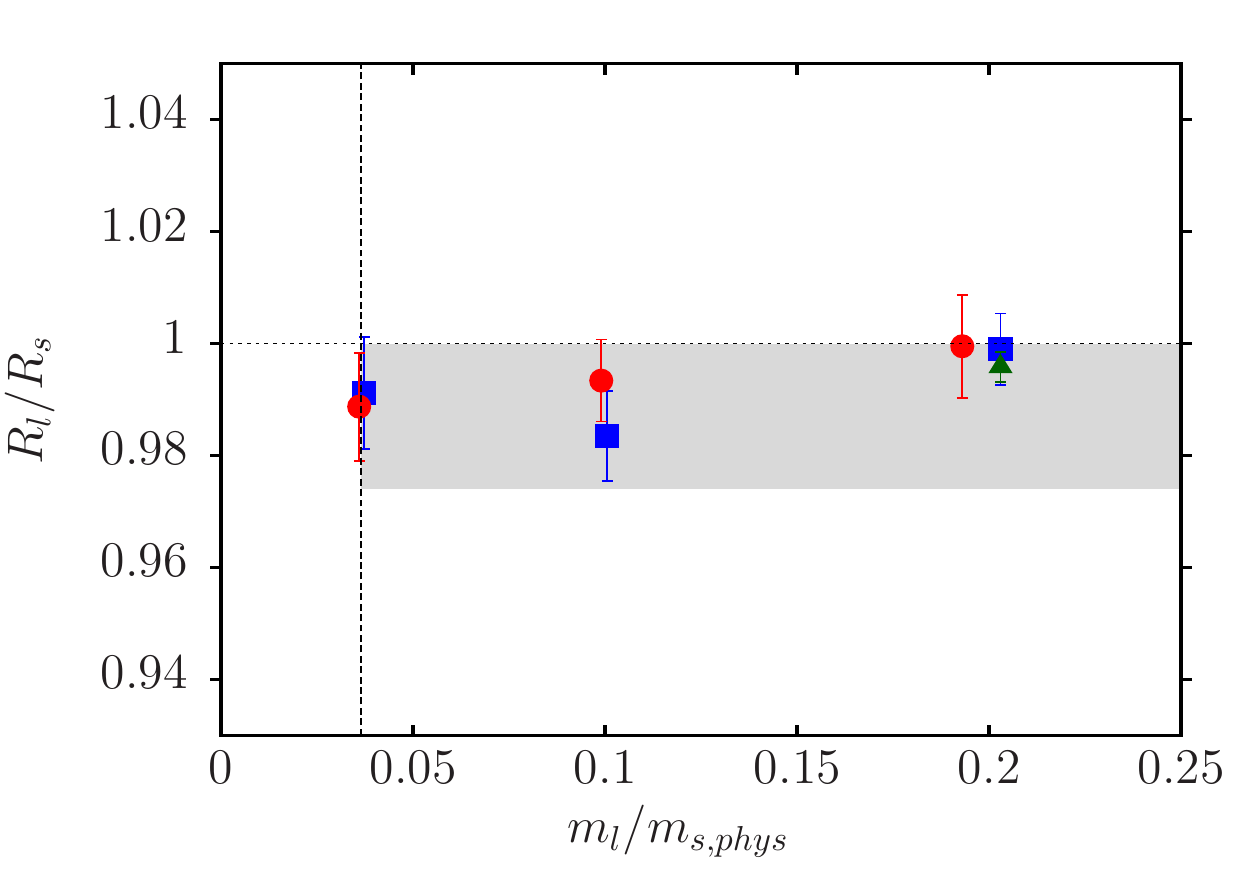}
\caption{Results for $R_l/R_s$, the SU(3)-breaking ratio 
of the ratio of vector to pseudoscalar decay constants 
(multiplied by the square root of the mass) plotted against 
the light quark mass (the average of $u$ and $d$)
in units of the physical $s$ quark mass. 
The filled blue squares gives results on very coarse 
lattices, red filled circles, on coarse and filled 
green triangles, on fine. 
The errors on the data points include statistical/fitting errors. 
The grey shaded band gives the physical result 
including all systematic errors discussed in the text.
The black dashed line shows the physical value of 
$m_l/m_{s,phys}$ and the blue dotted line indicates the value 1.0. 
}
\label{fig:BdBsrat}
\end{figure}

Figure~\ref{fig:BdBsrat} shows 
our results for $R_l/R_s$ on each ensemble, plotted 
against the light quark mass in units of the physical 
$s$ quark mass taken from~\cite{Dowdall:2011wh}. 
The results are very close to the value 1.0, but show a small 
downward trend as the light quark mass falls towards its 
physical value. There is no significant dependence on the 
lattice spacing. 

To fit the dependence of $R_l/R_s$ and extract a physical 
result, we use much of the 
same fit function as that given for $R_s$ in eq.~(\ref{eq:fitrs}). 
The two key differences are that the overall renormalisation factor 
now cancels, so that we can drop the factor $c\alpha_s^2$ from $F_2$, 
and that we now want to include a fitted dependence on $m_l$.  
We can also use the known similarity of $B_l$ and $B_s$ to constrain 
the fit further. For example, we know that decay constants for 
heavy-light and heavy-strange mesons differ by 
about 20\%~\cite{Dowdall:2013tga, oldfds}.  
This is in fact a very strong result, still 
true even when the light/strange quark is accompanied 
by a light or strange quark (see, for example,~\cite{Dowdall:2013rya} for 
$\pi$, $K$ and $\eta_s$ results). 
The $d\alpha_s$ term in eq.~(\ref{eq:fitrs}) takes account of 
missing radiative corrections 
to the sub-leading currents, $J^{(1)}$. We retain that term here but multiply its 
coefficient by 0.2 to allow for strange/light differences in 
the matrix element for $J^{(1)}$. We reduce the coefficient 
0.2 (allowed for the size of $\Lambda/m_b$) in front of 
discretisation errors and $m_b$ tuning terms by a 
further factor of 0.2 for the same reason.  
Finally, we include an additional term in the fit to allow for 
dependence on the light valence mass, since we have results 
for a variety of $m_l$ values. For this we include a term in 
$F_1$ of the form $h(m_l/(10m_{s,\mathrm{phys}}))$. The factor of 10 
once again is used to convert $m_s$ into the chiral scale of 1 GeV.  
The prior on $h$ is taken as 0.0(1.0). Since this term is 
already largely covered by including a term to allow for 
sea quark mass dependence, it has very little impact. 

The fit has a $\chi^2/{\mathrm{dof}}$ of 0.23
and gives a physical result:
\begin{equation}
\label{eq:rlrsres}
\frac{R_l}{R_s} = 0.987(13). 
\end{equation}
Since $R_l/R_s$ measures both SU(3)-breaking 
and spin-breaking effects in heavy-light meson 
decay constants we expect a result very close 
to 1.0. Our value is consistent with 1, but 
enables us to constrain any difference 
from 1.0 to a few percent. 
We will return 
to this in section~\ref{subsec:Bc} when comparing 
to results for $B_c$ mesons.   
A full error budget for $R_l/R_s$ is given in 
Table~\ref{tab:err}. 

Combining our result for $R_l/R_s$ with our earlier result 
for $R_s$ gives $R_l=0.945(26)$. 
Combining with the experimental 
value for the square root of the ratio of the meson masses, 
1.0043~\cite{pdg},
we obtain 
\begin{equation}
\label{eq:resrl}
\frac{f_{B^*_l}}{f_{B_l}} = 0.941(26)  
\end{equation} 
which is more than $2\sigma$ below 1.

\subsection{$f_{B_c}$ and $f_{B_c^*}$}
\label{subsec:Bc}
\begin{table}
\caption{Amplitudes for $J^{(0)}$ and $J^{(1)}$ for 
temporal axial and vector currents between the vacuum 
and the $B_c$ and $B_c^*$ mesons respectively, extracted 
from correlator fits and multiplied by $\sqrt{2}$ in accordance 
with eq.~(\ref{eq:phidef}). 
Results are in lattice units and the errors given are statistical/fit 
errors only. 
The ground-state energies determined from the fits agree with those 
given in~\cite{Dowdall:2012ab} and we do not repeat them here. 
\label{tab:phibc}
}
\begin{ruledtabular}
\begin{tabular}{lllll}
Set & $a^{3/2}\Phi_{B_c}^{(0)}$ & $a^{3/2}\Phi_{B_c}^{(1)}$ & $a^{3/2}\Phi_{B^*_c}^{(0)}$ & $a^{3/2}\Phi_{B^*_c}^{(1)}$ \\
\hline
1 & 0.83048(86) & -0.04792(5) & 0.8022(11) & 0.01541(3)\\
2 & 0.82001(45) & -0.04779(3) & 0.7904(6) & 0.01532(2) \\
\hline
4 & 0.58564(17) & -0.04068(2) & 0.54496(22) & 0.01267(1)\\
5 & 0.57350(11) & -0.04055(1) & 0.53195(14) & 0.01260(1)\\
\hline
7 & 0.36166(9) & -0.03158(1) & 0.31990(11) & 0.00941(1)\\
\end{tabular}
\end{ruledtabular}
\end{table}

\begin{table}
\caption{Coefficients $z_{A_0,c}$ and $z_1$, $z_2$
used in the matching factors to determine the 
decay constant for the $B_c$ meson (eq.~(\ref{eq:fimpbc})).  
$z_{A_0,c}$ is constructed from results 
given for the appropriate NRQCD bare masses and 
massive HISQ quarks with the appropriate 
values of $m_ca$ in~\cite{Monahan:2012dq} 
as $z_c = \eta_0-\tau_{10}$. 
$z_1$ and $z_2$ are results for massless HISQ 
quarks~\cite{Dowdall:2013tga, Monahan:2012dq}. 
The values of $\alpha_s$ used with these 
$z$ coefficients are given in Table~\ref{tab:params}.
\label{tab:zcharm}
}
\begin{ruledtabular}
\begin{tabular}{llll}
Set & $z_{A_0,c}$ & $z_1$ & $z_2$ \\
\hline
1 & -0.111(5)  & 0.024(3) & -1.108(4)\\
2 & -0.105(5)  & 0.024(3) & -1.083(4)\\
\hline
4 & -0.046(5) & 0.007(3) & -0.698(4)\\ 
5 & -0.041(5) & 0.007(3) & -0.690(4) \\ 
\hline
7 & -0.034(5) & -0.031(4) & -0.325(4)\\ 
\end{tabular}
\end{ruledtabular}
\end{table}

\begin{figure}[t]
\includegraphics[width=0.9\hsize]{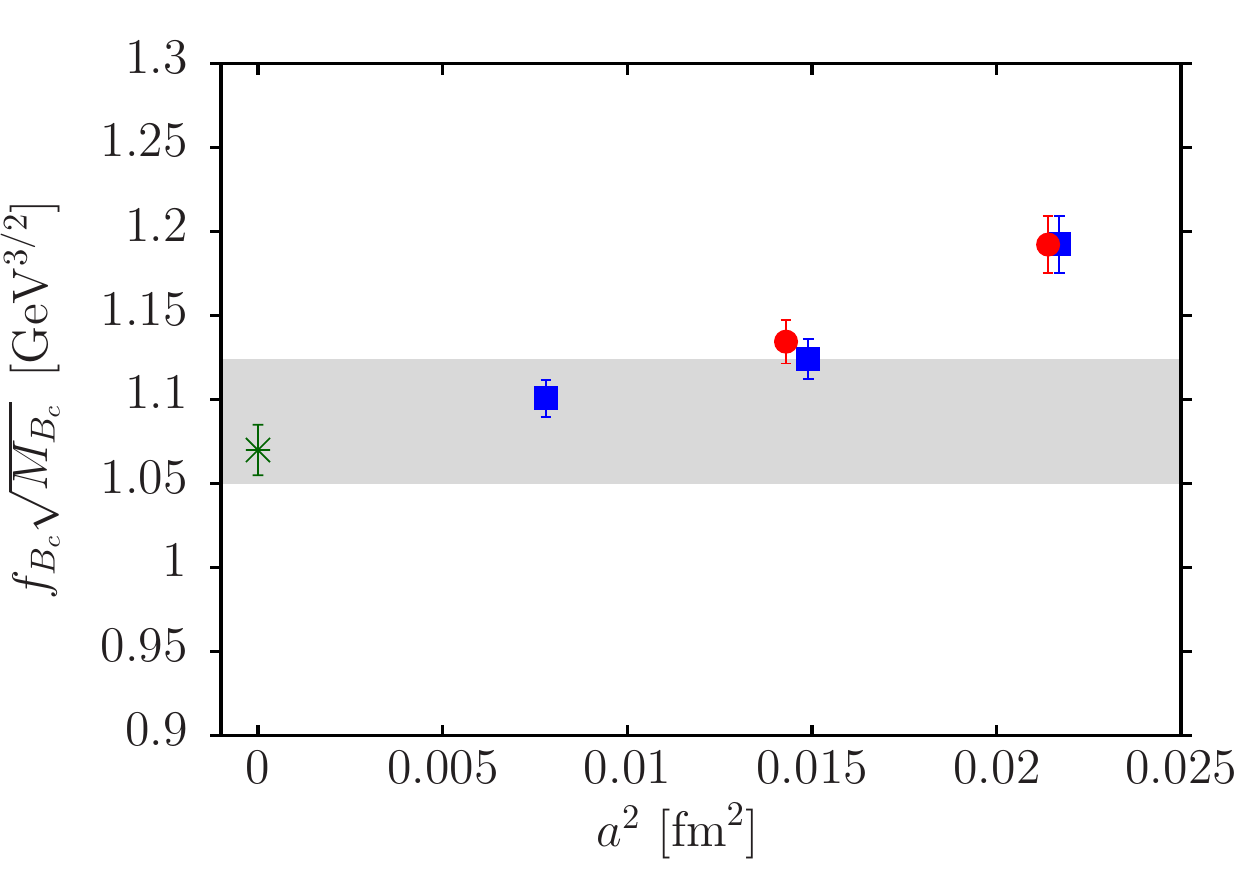}
\caption{
Results for the decay constant of 
the $B_c$ meson (multiplied by the square root of its 
mass) obtained with NRQCD $b$ quarks and HISQ $c$ quarks 
for ensembles at different values of the lattice spacing 
as described in the text. 
The errors on the points include uncertainties in the 
value of the lattice spacing and statistical/fitting errors. 
Blue filled squares give results at sea light quark 
mass $m_l/m_s=0.2$ and red filled squares at $m_l/m_s=0.1$. 
The grey shaded band gives the physical result 
including all systematic errors discussed in the text.
For comparison we include as the green burst the 
physical result obtained from using the HISQ formalism 
for both $b$ and $c$ quarks~\cite{McNeile:2012qf}. 
}
\label{fig:fbc}
\end{figure}

$B_c$ and $B^*_c$ meson correlation functions are 
calculated from NRQCD $b$ and HISQ $c$ propagators 
in exactly the same way as those described for 
NRQCD $b$ and HISQ $s$ or $l$ propagators in 
subsection~\ref{subsec:BBs}. We do not include the full 
set of ensembles used for the lighter HISQ quark 
mass calculations since experience has shown 
very little sea quark mass dependence for 
heavy meson correlators that do not include 
valence light quarks~\cite{Gregory:2010gm,Dowdall:2012ab}. 
We thus include ensembles at two different values of 
the sea $u/d$ quark mass for very coarse and coarse 
sets rather than three. 
The meson correlation functions are fit simultaneously 
so that correlations between them can be included in 
the determination of the ratio of amplitudes needed 
for the ratio of decay constants. 

The results for the matrix elements, $\Phi$, of
the leading, $J^{(0)}$, and subleading, $J^{(1)}$, pieces of the temporal 
axial and spatial vector currents are given in 
Table~\ref{tab:phibc}. We first discuss combining the 
results for the temporal axial current into a
value for the decay constant of the pseudoscalar 
$B_c$ meson. We will use a formula~\cite{Dowdall:2013tga} which is somewhat 
more accurate than that used in eq.~(\ref{eq:phidef}): 
\begin{eqnarray}
\label{eq:fimpbc}
f_{B_c}\sqrt{M_{B_c}} =  (1 + z_{A_0,c}\alpha_s)&\times&\left(\Phi_{B_c}^{(0)} + \Phi_{B_c}^{(1)} \right.\\
 &+& \left.z_1\alpha_s\Phi_{B_c}^{(1)}+ z_2\alpha_s \Phi_{B_c}^{(2)}\right) .\nonumber 
\end{eqnarray}
$z_1\alpha_s$ is an additional radiative correction to the sub-leading 
current $J^{(1)}$. $z_2\alpha_s$ multiplies an additional sub-leading 
current which has the same matrix element as $J^{(1)}$ and so does not 
need to be separately calculated. The $z$ coefficients now have to 
be calculated for massive HISQ quarks with a mass in lattice units 
corresponding to our values for $am_c$ on the different ensembles. 
This has been done for $z_{A_0,c}$ and the values are given in 
Table~\ref{tab:zcharm}. They differ slightly from those for massless HISQ 
quarks in 
Table~\ref{tab:zlight} but are still very much less than 1. 
The $z_1$ and $z_2$ coefficients have only been calculated 
for massless HISQ quarks and these are also given in Table~\ref{tab:zcharm}. 
There is then a systematic error in our formula of eq.~(\ref{eq:fimpbc}) 
as a result of using the massless $z_1$ and $z_2$ coefficients and 
we will allow for that in our error budget along with systematic 
errors from unknown higher order terms in the overall renormalisation factor. 

The results obtained from applying eq.~(\ref{eq:fimpbc}) are plotted 
in Figure~\ref{fig:fbc} as a function of lattice 
spacing. We see, as expected, very little change between 
ensembles with similar lattice spacings but 
different sea $u/d$ quark masses. 
To determine a physical value for the decay constant 
we fit the results to a functional form that includes 
allowance for systematic errors in the lattice QCD 
calculation. 

The systematic errors have the same sources as those discussed 
for $R_s$ in section~\ref{subsec:BBs} and we will use the same 
fit form as that given in eq.~(\ref{eq:fitrs}) and we reproduce 
that below as eq.~(\ref{eq:fitbc}) with the modifications appropriate here. As in 
section~\ref{subsec:BBs}, the major source of uncertainty here 
comes from missing higher order terms in the matching 
of the NRQCD-HISQ current to continuum QCD. 
This is taken account of in eq.~(\ref{eq:fitbc}), as before, by 
the term $F_2(\alpha_s)$ which includes an $\alpha_s^2$ term with 
coefficient $c$ in the overall renormalisation factor and a 
term with coefficient $d$ that allows for systematic errors 
in the $\alpha_s$ corrections to the $J^{(1)}$ current contribution 
included in eq.~(\ref{eq:fimpbc}) from the fact that $z_1$ and 
$z_2$ are taken for massless HISQ quarks. Given the values we 
have for $z_{A_0,c}$ and the dependence on $am_c$ seen in that 
coefficient, we do not expect coefficients $c$ and $d$ to be 
large and we take priors on their fit values of 0.0(0.2). 

From Fig.~\ref{fig:fbc} we see significant lattice spacing dependence 
in the results and we must allow both for regular lattice spacing 
dependence and that coming from the NRQCD action. 
This dependence is included in factor $F_1$. 
The regular lattice spacing coming the HISQ action 
can have a scale set by $m_c$ in this case and we expect that to 
dominate. We take $m_c$ to be 1 GeV here. 
The analysis of discretisation errors for $c$ quarks in the HISQ 
action~\cite{Follana:2006rc} shows that the dependence comes from 
terms suppressed by powers of the velocity of the $c$ quark.  
Since $v_c^2 \approx 0.5$ in a $B_c$~\cite{Gregory:2009hq} we 
include a factor of 0.5 in front of the terms allowing 
for discretisation errors. 
We must also allow for dependence on the $u/d$ quark mass in the 
sea, as before, and for mistuning of the $b$ quark mass. 
For mistuning of the $b$ quark mass we allow a conservative factor of 0.3 based 
on the variation in decay constants between 
heavyonium mesons (see Fig.~\ref{fig:vpsrat}). 

Our fit function is: 
\begin{eqnarray}
\label{eq:fitbc}
f_{B_c}\sqrt{M_{B_c}}(a,m) &=& \left(f_{B_c}\sqrt{M_{B_c}}\right)_{\mathrm{phys}} \\
&\times& F_1(a,m) /F_2(\alpha_s) ;\nonumber \\
F_2(\alpha_s) &=& (1 + c\alpha_s^2 + 0.08d\alpha_s) \nonumber \\
F_1(a,m) &=& 1 + 0.5\sum_j e_j (m_c a)^{2j} \nonumber \\  
&+& 0.3 f \frac{\delta m_b}{m_{b,\mathrm{phys}}} + g \frac{\delta m_{\mathrm sea}}{10m_{\mathrm{sea,phys}}}  \nonumber \\
c &=& c_1\times(1+c_2\delta x_m + c_3 (\delta x_m)^2) \nonumber \\
d &=& d_1\times(1+d_2\delta x_m + d_3 (\delta x_m)^2) \nonumber \\
e_1 &=& e_{11}\times(1+e_{12}\delta x_m + e_{13} (\delta x_m)^2) \nonumber 
\end{eqnarray}
We take prior values on all coefficients to be 0.0(1.0) except for 
the physical value on which we take 1.0(2), $c$ and $d$, on which we take 0.0(2)
and $e_1$ on which we take 0.0(3) (since it is $\mathcal{O}(\alpha_s)$). 
The error on the plotted values in Figure~\ref{fig:fbc} is dominated 
by the uncertainty in the value of the lattice 
spacing (given in Table~\ref{tab:params}). In doing the fit we allow 
for half of this error to be correlated between ensembles 
(since it comes from systematic uncertainties from the 
NRQCD calculation used to fix the lattice spacing~\cite{Dowdall:2011wh}) 
and half to be uncorrelated.  

The fit gives a $\chi^2/\mathrm{dof}$ of 0.11 
and a physical value for $f\sqrt{M}$ for the $B_c$ 
of 1.087(37) $(\mathrm{GeV})^{3/2}$. The 3.4\% 
uncertainty is split between 1.2\% from matching 
and 3.2\% from other sources, dominated by 
lattice spacing uncertainties and 
discretisation errors.  
We have checked that missing out $z_1$ and $z_2$ 
from eq.~(\ref{eq:fimpbc}) and allowing a larger prior 
of 0.0(1.0) on the coefficient $d$ in eq.~(\ref{eq:fitbc}) 
gives a physical result with almost the same central 
value and uncertainty. 

Our physical result is plotted as a grey band in 
Figure~\ref{fig:fbc}. It agrees very well with 
our result of 1.070(15) ${\mathrm{GeV}}^{3/2}$~\cite{McNeile:2012qf} 
based on using the HISQ action for 
a heavy quark combined with a HISQ $c$ quark and working at 
a range of heavy quark masses between $c$ and $b$   
on lattices with a range of lattice spacings from 
0.15 fm down to 0.045 fm. The HISQ-HISQ result has an 
uncertainty which is a factor of 2 smaller than the NRQCD-HISQ 
result we give here. This is because the HISQ-HISQ 
current is absolutely normalised in the calculation of 
pseudoscalar decay constants and the calculation is done 
over a wider range of values of the lattice spacing for 
better control of discretisation errors. Good agreement 
between the HISQ-HISQ result and NRQCD-HISQ result 
was already seen for the $B_s$ in~\cite{McNeile:2011ng, Dowdall:2013tga}
and this further test increases our confidence in 
our handling of lattice QCD errors. In particular it is an 
important test of our normalisation of improved NRQCD-HISQ 
currents that are also in use for semileptonic decay rate 
calculations underway on these gluon field configurations.  

Using the experimental value of the $B_c$ meson mass
of 6.276(1) GeV~\cite{Aaij:2013gia} we can convert our 
value for $f\sqrt{M}$ into a result for the decay 
constant:
\begin{equation}
\label{eq:fbcval}
f_{B_c} = 0.434(15) \mathrm{GeV} .
\end{equation}
Again this agrees with our earlier result using 
HISQ quarks of 0.427(6) GeV~\cite{McNeile:2012qf}.

\begin{table}
\caption{Column 2 gives the Coefficient $z_{V_i,c}$ 
needed for the one-loop renormalisation factor 
for the vector decay constant 
$B^*_c$. 
$z_{V_i,c}$ is constructed from results 
given for the appropriate NRQCD bare masses and 
massive HISQ quarks with the appropriate 
values of $m_ca$ in~\cite{Monahan:2012dq} 
as $z_c = \eta_0-\tau_{10}$. 
Column 3 gives 
$\delta z_c$, which is the difference 
between $z_{V_i,c}$ and $z_{A_0,c}$ from 
Table~\ref{tab:zcharm}. 
Column 4 gives the difference between $\delta z_c$ 
and the 
corresponding value $\delta z$ for massless 
HISQ quarks (from Table~\ref{tab:zlight}).  
The values of $\alpha_s$ used with these 
$z$ coefficients are given in Table~\ref{tab:params}.
Column 5 gives the unrenormalised ratio of vector to pseudoscalar 
amplitudes (see text) determined from the simultaneous fit 
to $B^*_c$ and $B_c$ meson correlators. 
Results from columns 4 and 5 are 
 used in the determination of $R_c/R_s$. 
\label{tab:deltazcharm}
}
\begin{ruledtabular}
\begin{tabular}{lllll}
Set & $z_{V_i,c}$ & $\delta z_c$ & $\delta z_c - \delta z$ & $R_c^{\mathrm{unren.}}$ \\
\hline
1 & -0.166(5)  & -0.055(7)  & 0.047(8) & 1.0447(14)\\
2 & -0.160(5)  &  -0.055(7)  & 0.044(8) & 1.0434(8)\\
\hline
4 & -0.073(5) & -0.027(7) & 0.052(8) & 1.02324(27)\\ 
5 & -0.068(5) &  -0.027(7) & 0.046(8) & 1.02175(17) \\ 
\hline
7 & -0.013(5) & 0.021(7) &  0.058(8) & 0.99766(22)\\ 
\end{tabular}
\end{ruledtabular}
\end{table}

\begin{figure}[t]
\includegraphics[width=0.9\hsize]{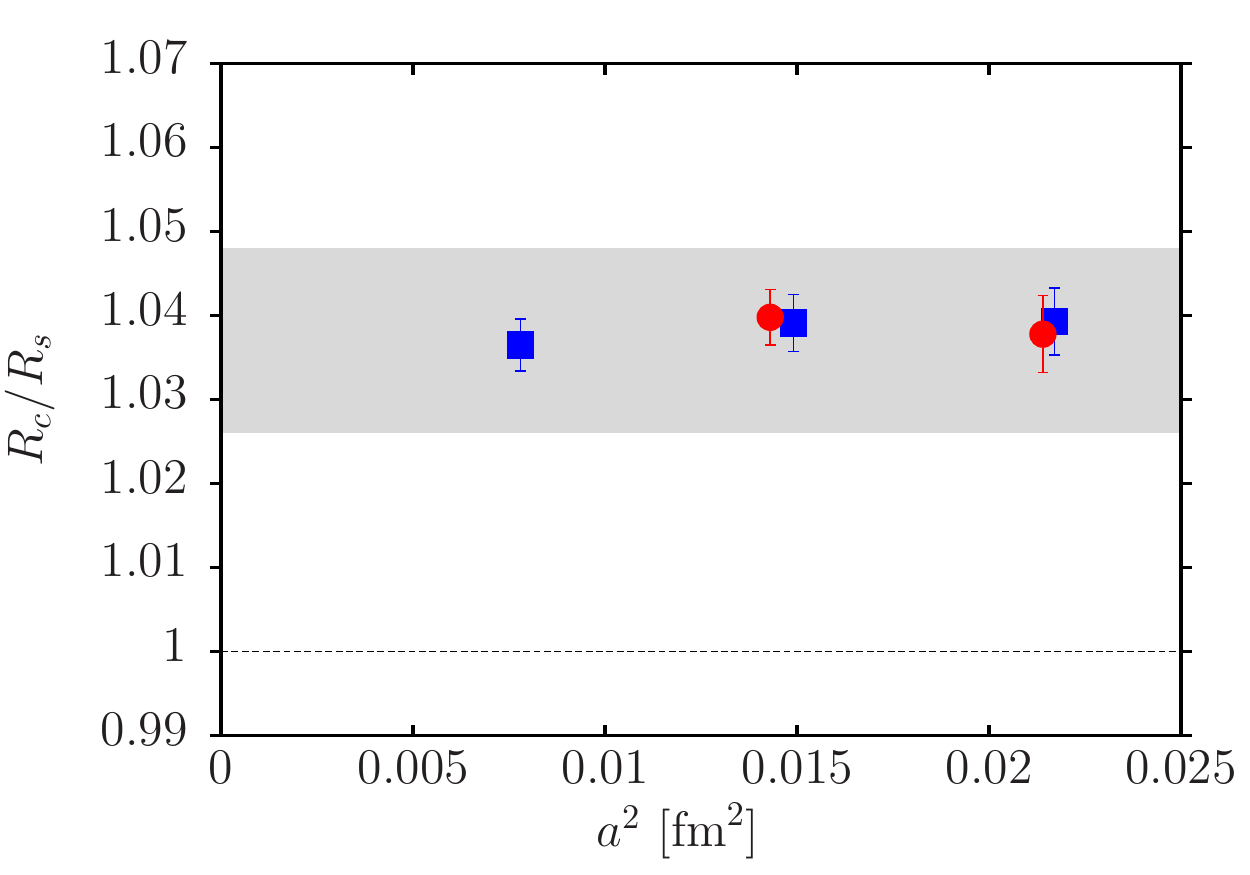}
\caption{
Results for the ratio of $R_c$ to $R_s$ plotted 
against the square of the lattice spacing. 
$R_c$ is the ratio of vector to pseudoscalar 
decay constants ($f\sqrt{M}$) for the $B^*_c$ and 
$B_c$ and $R_s$ is the corresponding ratio for 
the $B^*_s/B_s$. Filled blue squares are results on 
ensembles with $m_l/m_s=0.2$ and filled red circles
results on ensembles with $m_l/m_s=0.1$. 
The errors on the points are statistical errors only 
(including those from $(\delta z_c - \delta z)$). 
The grey shaded band gives the physical result 
including all systematic errors as discussed in the text.
The black dashed line marks the value 1.0. 
}
\label{fig:bcbsrat}
\end{figure}

The vector to pseudoscalar decay constant ratio, $R_c$, 
is obtained in an analogous way to that for the 
$B_s$ and $B^*_s$ mesons described in Section~\ref{subsec:BBs}. 
The formula we use is that given in eq.~\ref{eq:frat}, in 
which $z_{A_0}$ and $z_{V_i}$ are the coefficients calculated 
for temporal axial and spatial vector currents respectively 
using HISQ quark mass values appropriate to $c$ quarks. 
Values for $z_{A_0,c}$ are given in Table~\ref{tab:zcharm} 
and values for $z_{V_i,c}$ are given in Table~\ref{tab:deltazcharm}. 
Table~\ref{tab:deltazcharm} also gives $\delta z_c$, the 
difference between the two, which is needed for the 
decay constant ratio in eq.~\ref{eq:frat}. Note that we 
are now neglecting radiative corrections to the 
current correction $J^{(1)}$ (i.e. the terms with coefficients 
$z_1$ and $z_2$ in eq.~(\ref{eq:fimpbc})) since we do not 
have these terms for the vector current. 

Table~\ref{tab:phibc} gives the results for the amplitudes 
that we need to construct the decay constants and their 
ratio. We see that the qualitative features 
of the results are the same i.e. that the amplitude of the 
leading order current is smaller for the vector than for 
the pseudoscalar meson, lowering 
the vector to pseudoscalar decay constant ratio, whereas the 
current correction contributions have opposite effect. 
The impact of the current corrections is a few percent less 
than in the $B_s$ case but varies more strongly with lattice 
spacing. 

In a similar approach to that used for $R_l$ in 
Section~\ref{subsec:BBs} we will 
study $R_c$ through its ratio with $R_s$. In this case the 
renormalisation factor does not cancel completely at 
$\mathcal{O}(\alpha_s)$ since $\delta z_c$ is not equal 
to $\delta z$. Instead we have a renormalisation factor 
for $R_c/R_s$ which is $(1+[\delta z_c - \delta z]\alpha_s)$, 
i.e. we can write: 
\begin{equation}
\label{eq:rcrsdef}
\frac{R_c}{R_s} = (1+[\delta z_c - \delta z]\alpha_s) \left(\frac{\Phi^{(0)}_{B^*_c}+\Phi^{(1)}_{B^*_c}}{\Phi^{(0)}_{B_c}+\Phi^{(1)}_{B_c}}\right)\left(\frac{\Phi^{(0)}_{B_s}+\Phi^{(1)}_{B_s}}{\Phi^{(0)}_{B^*_s}+\Phi^{(1)}_{B^*_s}}\right)
\end{equation} 
Values of $(\delta z_c - \delta z)$ are given in 
Table~\ref{tab:deltazcharm}. We see that these are small and 
independent of the value of $am_b$ within statistical 
uncertainties. Since the dependence 
on $am_b$ comes from the NRQCD action it is not surprising 
to find some cancellation between these two cases. 
The remaining small 
renormalisation then reflects the fact that 
the $c$ quark mass is not zero (i.e. $m_c/m_b \ne 0$). 
Table~\ref{tab:deltazcharm} gives in the final column results 
for the appropriate ratio of sums of amplitudes needed in 
eq.~(\ref{eq:rcrsdef}), i.e. $R^{\mathrm{unren.}}_c$. This 
can be combined with $R^{\mathrm{unren.}}_s$ from Table~\ref{tab:phibs} 
and the small renormalisation applied to form $R_c/R_s$. 

Figure~\ref{fig:bcbsrat} gives results for the ratio of 
$R_c$ to $R_s$ from eq.~(\ref{eq:rcrsdef}) as a function of lattice spacing. We 
see that the results are independent of lattice spacing and 
sea quark mass. Importantly the values obtained are all significantly 
larger than 1.0,
showing that the vector to pseudoscalar decay constant 
ratio is sensitive to the mass 
of the light quark combined with the $b$.  
Half of the difference from 1.0 comes from the raw amplitudes 
and the other half from the renormalisation factor in 
eq.~(\ref{eq:rcrsdef}). 

In fitting this ratio as a function of lattice spacing to 
obtain a physical result we will use the same fit form 
as that used in subsection~\ref{subsec:BBs}, eq.~(\ref{eq:fitrs}). 
The only change in form that we make is to remove $am_b$ dependence 
from the coefficient of unknown $\alpha_s^2$ renormalisation terms in 
the factor $F_2$ assuming 
they follow the same form as discussed above for the $\mathcal{O}(\alpha_s)$ 
term. We take the radiative correction terms for the $J^{(1)}$ currents in $F_2$ 
to have the same form as in eq.~(\ref{eq:fitrs}) but allow for cancellation 
between $R_c$ and $R_s$ by giving that term coefficient 0.03 rather 
than 0.08. For the discretisation errors and $m_b$ tuning error 
terms in $F_1$ we likewise allow for cancellation between $R_c$ and 
$R_s$ by giving these terms coefficient 0.1 rather than 0.2. 

The fit to our results gives $\chi^2/{\mathrm{dof}}$ of 0.1 and 
a physical result of 
\begin{equation}
\label{eq:rcrsres}
\frac{R_c}{R_s} = 1.037(12) 
\end{equation}
where we have allowed a 0.5\% 
uncertainty from missing higher-order relativistic current corrections.  
This value is $3\sigma$ greater than 1.0 giving a clear indication 
that $R_q$ increases as the mass of the quark $q$ increases. This 
is consistent with what was found (with much lower significance)
in Section~\ref{subsec:BBs} for $q = l$ and $s$. 
Our physical result for $R_c/R_s$ is plotted as the 
grey band in Figure~\ref{fig:bcbsrat}. 
A full error budget for $R_c/R_s$ is given in Table~\ref{tab:err}. 

Using our earlier value for $R_s$ of 0.957(23) we obtain
$R_c$ = 0.992(27).  
We convert $R_c$ into a ratio of the decay constants of the $B^*_c$ and 
$B_c$ mesons by dividing by the square root of the ratio of the masses. 
For this we use the experimental value for the $B_c$ 
mass of 6.276(1) GeV~\cite{Aaij:2013gia} 
and our lattice QCD result for the mass difference between 
$B^*_c$ and $B_c$ of 54(3) MeV~\cite{Dowdall:2012ab}. This gives a 
mass ratio for $B^*_c$ to $B_c$ of 1.0086(5).  We then obtain 
\begin{equation}
\label{eq:rcres}
\frac{f_{B^*_c}}{f_{B_c}} = 0.988(27) . 
\end{equation}

\subsection{Heavy Quark Mass Dependence}
\label{subsec:mQ}
\begin{table*}
\caption{
The coefficients $c_1$, $c_5$, $c_4$ used 
in the NRQCD action (eq.~(\ref{eq:Hamiltonian})) for values of 
the heavy quark mass in lattice units given in column 2. 
$c_6$ is equal to $c_1$ and $c_2$ and $c_3$ are set to 1.0.
Amplitudes for $J^{(0)}$ and $J^{(1)}$ for 
temporal axial and vector currents between the vacuum 
and heavy pseudoscalar and vector mesons respectively, 
made from a heavy NRQCD quark and a HISQ $s$ quark 
and denoted $H_s$ and $H^*_s$. 
The different heavy quark masses in lattice units used on sets 1 and 4 
are given in column 2. 
Results are in lattice units and the errors given are statistical/fit 
errors only. 
\label{tab:lighterb}
}
\begin{ruledtabular}
\begin{tabular}{llllllllllll}
Set & $am_h$ & $c_1$ & $c_5$ & $c_4$ & $z_1$ & $z_2$ & $a^{3/2}\Phi_{H_s}^{(0)}$ & $a^{3/2}\Phi_{H_s}^{(1)}$ & $a^{3/2}\Phi_{H^*_s}^{(0)}$ & $a^{3/2}\Phi_{H^*_s}^{(1)}$ & $R_s$ \\
\hline
1 & 1.91 & 1.29 & 1.18 & 1.19 & -0.031(4) & -0.325(4) & 0.3401(6) & -0.04274(10) & 0.2958(14) & 0.01273(8) & 1.0242(43)\\
 & 2.66  & 1.36 & 1.19 & 1.21 & 0.007(3) & -0.698(4) & 0.3600(8) & -0.03430(9) &  0.3233(16) & 0.01052(7) & 0.9969(41)\\
\hline
4 & 1.91 & 1.26 & 1.15 & 1.18 & -0.031(4) & -0.325(4) & 0.2602(4) & -0.02958(6) & 0.2237(6) & 0.00865(3) & 0.9956(21)\\
 & 3.297 & 1.31 & 1.17 & 1.21 & 0.024(3) & -1.108(4) & 0.2802(7) & -0.01997(6) & 0.2534(8) & 0.00611(2) & 0.9657(21)\\
\end{tabular}
\end{ruledtabular}
\end{table*}

\begin{figure}[t]
\includegraphics[width=0.9\hsize]{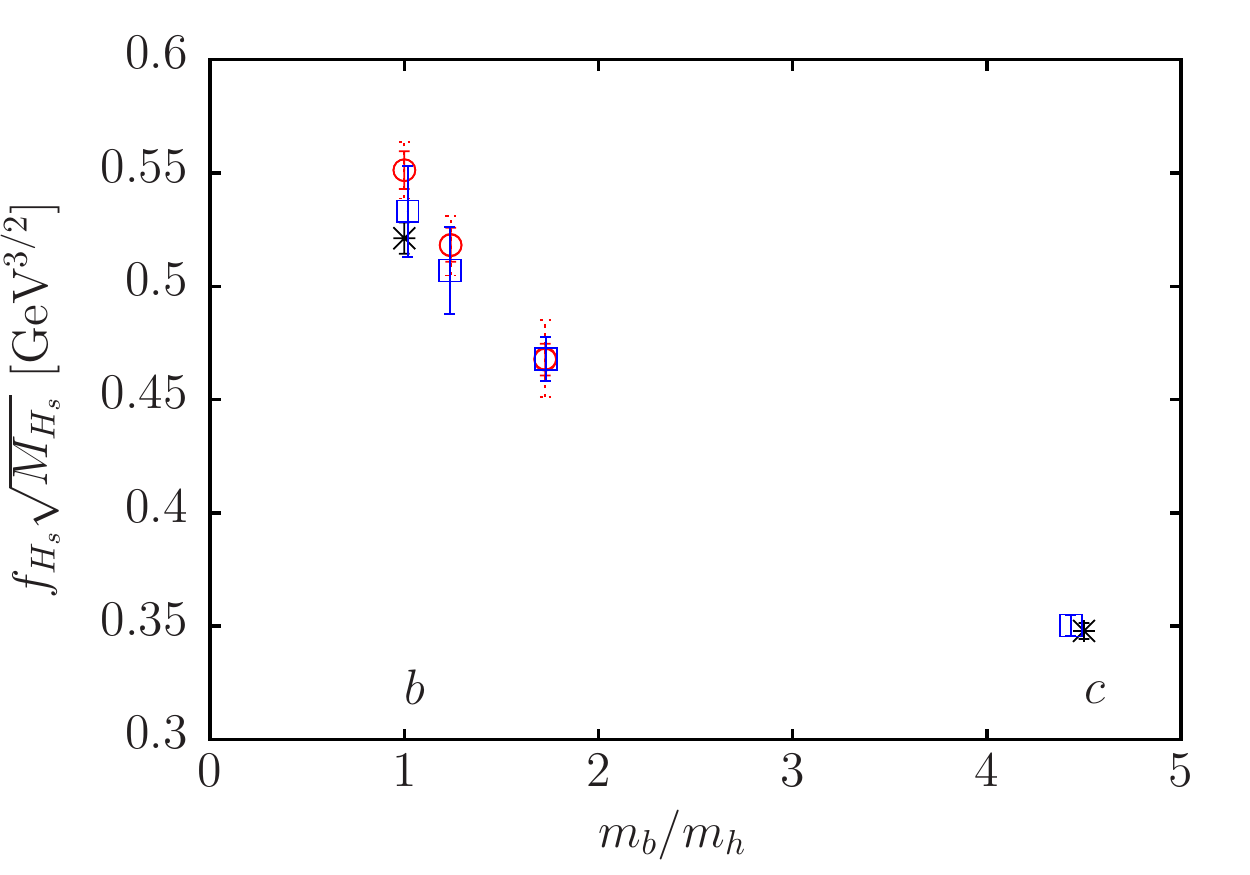}
\caption{ 
Results for the decay constant of the pseudoscalar 
heavy-strange meson $H_s$ multiplied by the square 
root of its mass as a function of the heavy quark 
mass in units of the physical $b$ quark mass. 
Open red circles are results on set 1
ensembles from improved NRQCD heavy quarks combined 
with HISQ $s$ quarks (from Tables~\ref{tab:phibs} and~\ref{tab:lighterb}). 
The open blue squares are results from $a=$ 0.044 fm lattices
using the HISQ formalism for both $b$ and $s$~\cite{McNeile:2011ng}. 
The solid error bars on both sets of points include statistics and the (correlated) 
uncertainty in the value of the lattice spacing. The dotted 
error bars on the NRQCD points include in addition an estimate of NRQCD systematic 
errors~\cite{Dowdall:2013tga}.  
The black bursts are the final physical values for 
the $D_s$ and $B_s$~\cite{newfds, Dowdall:2013tga}. 
}
\label{fig:nrhisqf}
\end{figure}

\begin{figure}[t]
\includegraphics[width=0.9\hsize]{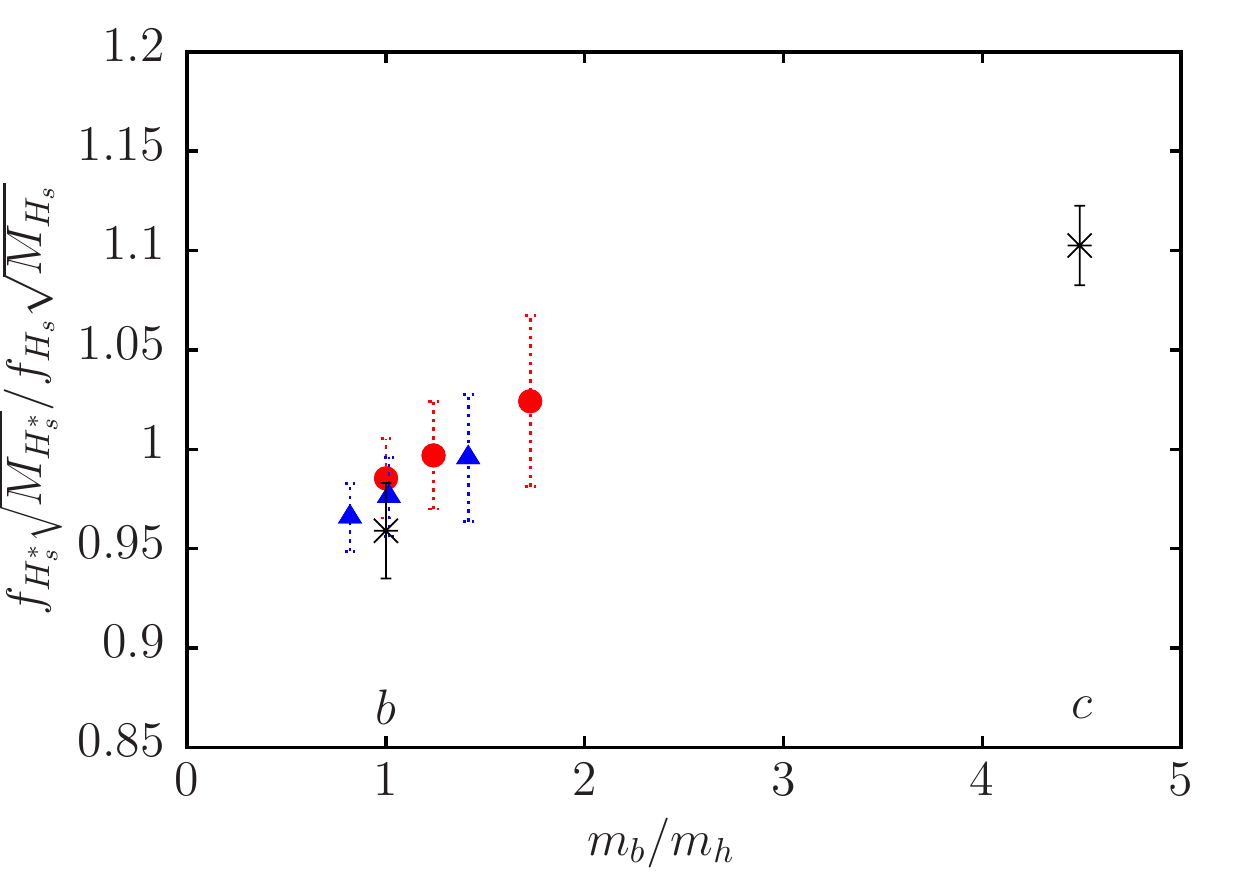}
\caption{ 
Results for the ratio of vector to pseudoscalar decay 
constants for heavy-strange mesons made with a range 
of heavy quark masses, $m_h$, as a function of the 
heavy quark mass in units of $m_b$. 
Filled red circles give results from very coarse set 1, 
and filled blue triangles from coarse set 4, for NRQCD 
heavy quarks (Tables~\ref{tab:compbs} and~\ref{tab:lighterb}). 
Dotted error bars include an estimate of NRQCD systematic 
errors. 
Black bursts indicate the physical result for $B^*_s/B_s$ mesons 
from this paper and for $D^*_s/D_s$ mesons using HISQ $c$ and 
$s$ quarks from~\cite{Donald:2013sra}. 
}
\label{fig:hrat}
\end{figure}

In this subsection we give results for calculations that use 
NRQCD quarks with masses lighter than that of the $b$ in order 
to study the heavy-quark mass dependence of decay constants 
and their ratios and make a link 
between $b$ and $c$. Using the HISQ action 
we have previously mapped out the dependence of pseudoscalar 
decay constants and quark 
masses~\cite{McNeile:2010ji, McNeile:2011ng, McNeile:2012qf, Chakraborty:2014zma} 
in this region in some detail, and we will be able to compare 
to these results. 

The HISQ 
action has the smallest discretisation errors of any quark 
action in current use, since it removes tree-level $a^2$ errors 
and has no odd powers of $a$ appearing. It is therefore a very 
good action for $c$ physics~\cite{Follana:2006rc, newfds, Donald:2012ga}. 
Raising the mass from that of $c$ requires fine lattices to keep 
masses in lattice units below $ma=1$, where, naively, it 
might be expected that discretisation errors would become large. 
It is possible to reach the $b$ on 
`ultrafine' lattices with a lattice spacing as small as 
$a=0.045$ fm. 
This has given accurate results for $m_b$, $f_{B_s}$ 
and $f_{B_c}$~\cite{McNeile:2010ji, McNeile:2011ng, McNeile:2012qf} because 
we can use operators that are absolutely normalised. 

For NRQCD the issues are complementary ones. In this case 
we have systematic control of a 
non-relativistic effective theory. Discretisation errors are much 
smaller, having a scale set by internal momenta rather than the quark 
mass. In this case naive arguments suggest that we need $ma > 1$ 
to control coefficients of relativistic correction operators, for 
high precision. 
In fact for $b$ quarks on the ensembles we use here, with
lattice spacing values ranging from 0.15 fm down to 0.09 fm, 
values of $ma$ are well above 1 and there is significant headroom 
to reduce the mass, particularly on the coarser 
lattices. Since the ratio of $c$ 
to $b$ quark mass is 4.5~\cite{McNeile:2010ji}, we cannot reach 
the $c$ quark mass with $ma>1$ even on the very coarse lattices. 
However, it is still of interest to vary the mass and compare the 
mass-dependence using NRQCD heavy quarks to that 
obtained from a completely different perspective, in terms 
of systematic errors, using HISQ quarks.   

We have already shown that using HISQ $b$ quarks and 
NRQCD $b$ quarks gives results in agreement for the 
decay constant of the $B_s$~\cite{McNeile:2011ng, Dowdall:2013tga} 
and the $B_c$ (~\cite{McNeile:2012qf} and subsection~\ref{subsec:Bc}). 
Here we will illustrate how well this agreement continues to lighter 
masses.  

We work on one ensemble each from 
the very coarse (set 1) and coarse (set 4) lattices. We will focus on results 
using $s$ HISQ quarks where, as we have seen, dependence 
on the sea $u/d$ mass is negligible.  
It is most convenient to use the same values of $am_b$ 
as those used before on the finer lattices, since then 
the coefficients of the radiative corrections to terms 
in the NRQCD Hamiltonian are already known. 
We simply have to change the value of $\alpha_s$ multiplying 
them on the coarser lattices. 
In Table~\ref{tab:lighterb} 
we give the coefficients that we use for heavy quark masses 
$am_h=1.91$ and 2.66 on very coarse set 1 and for 
$am_h=1.91$ and 3.297 on coarse set 4. On very coarse set 
1 the lightest $am_h$ then corresponds to 1.91/3.297= 0.58 times 
$m_b$. On coarse set 4 the mass 3.297 is higher than $m_b$ (since 
there $am_b=2.66$, see Table~\ref{tab:upsparams}), but 
$am=1.91$ corresponds to 0.72 times $m_b$.  
The coefficients are calculated by combining the one-loop 
coefficients at the appropriate $am_b$ values 
given in~\cite{Dowdall:2011wh, Hammant:2011bt} with the appropriate $\alpha_s$ value 
(also given in~\cite{Dowdall:2011wh}) for that lattice spacing. 
These coefficients are then used in the NRQCD action 
(eq.~(\ref{eq:Hamiltonian})) along with relevant tadpole-improvement 
factors given in Table~\ref{tab:upsparams} for that ensemble. 

We again use a local and two smeared sources for the 
NRQCD propagators, with smearing radii as 
given in Table~\ref{tab:upsparams}. 
We combine the NRQCD propagators with those for the 
HISQ $s$ quarks on each ensemble. Table~\ref{tab:lighterb} 
gives results for the amplitudes for the leading-order and 
relativistic correction currents for the heavy-strange 
pseudoscalar meson ($H_s$) and vector meson ($H^*_s$). 
These are obtained from simultaneous fits to the vector 
and pseudoscalar meson correlators as described in 
subsection~\ref{subsec:2pt}. 

To determine the pseudoscalar decay constant, $f_{H_s}$, 
we are able to use a more accurate formula than the one given in
eq.~(\ref{eq:phidef}), because additional current corrections 
coefficients are available in this case (only). 
We can use the formula accurate 
through $\alpha_s\Lambda/m_h$ given in~\cite{Dowdall:2013tga}: 
\begin{eqnarray}
\label{eq:fimp}
f_H\sqrt{M_H} =  (1 + z_{A_0}\alpha_s)&\times&\left(\Phi_{A_0}^{(0)} + \Phi_{A_0}^{(1)} \right.\\
 &+& \left.z_1\alpha_s\Phi_{A_0}^{(1)}+ z_2\alpha_s \Phi_{A_0}^{(2)}\right) .\nonumber 
\end{eqnarray}
$z_1\alpha_s$ is an additional radiative correction to the sub-leading 
current $J^{(1)}$. $z_2\alpha_s$ multiplies an additional sub-leading 
current which has the same matrix element as $J^{(1)}$ and so does not 
need to be separately calculated. The coefficients $z_1$ and $z_2$ are 
given for the masses we use in Table~\ref{tab:lighterb}. The $z_{A_0}$ 
values are in Table~\ref{tab:zlight} and $\alpha_s$ values in Table~\ref{tab:params}.  

Figure~\ref{fig:nrhisqf} shows results for $f_{H_s}\sqrt{M_{H_s}}$ 
as a function of inverse heavy quark mass in units of the physical 
$b$ quark mass. The results for set 1 are shown as open red circles
including the value at the $b$ quark mass ($am_b=3.297$) from Table~\ref{tab:phibs} as well as the results for lighter heavy quark masses from 
Table~\ref{tab:lighterb}. The solid error bar is the dominant error in the 
raw results coming from the uncertainty in the lattice spacing. The dotted 
error bar includes an estimate of systematic errors from NRQCD coming 
from missing $\alpha_s^2$ renormalisation and $(\Lambda/m_h)^2$ current 
corrections. The latter systematic error grows as $m_h$ falls.   
The open blue squares give results from `ultrafine' ($a=$0.044fm) 
$n_f=2+1$ lattices using the HISQ formalism for the 
heavy quark~\cite{McNeile:2011ng}. 
These results were part of an analysis of the heavy-strange pseudoscalar 
meson decay constant that spanned the range from $c$ to $b$. 

The plot shows good consistency between the two sets of results, which 
use very different formalisms on lattices that differ in lattice spacing 
by over a factor of 3. The black stars mark the final physical result 
for the $B_s$~\cite{Dowdall:2013tga} and $D_s$~\cite{newfds} decay 
constants obtained by HPQCD after performing 
a fit including discretisation uncertainties.  

Table~\ref{tab:lighterb} also includes results for the vector 
to pseudoscalar ratio of decay constants, $R_s = f_{H^*_s}\sqrt{M_{H^*_s}}/f_{H_s}\sqrt{M_{H_s}}$. 
This is defined from eq.~(\ref{eq:frat}) up to missing 
$\alpha_s^2$ and $\alpha_s\Lambda/m_b$ matching uncertainties. 
These are plotted as a function of the inverse heavy quark 
mass in units of the $b$ quark mass in Figure~\ref{fig:hrat}, 
including also results from Table~\ref{tab:compbs} at the $b$ quark 
mass. We see, as expected, that the values rise as $m_b/m_h$ grows 
towards the $c$ quark mass. The dotted error bars include an estimate 
of the (correlated) systematic error from missing factors in the 
matching of the NRQCD current to full QCD. These are estimated by 
rescaling results from our study here for the $B_s$ (subsection~\ref{subsec:BBs}).   
The missing terms are: $\alpha_s^2$ terms in the overall renormalisation which 
are taken to be independent of $m_h$;  $\alpha_s\Lambda/m_h$ 
current corrections which grow linearly with $m_b/m_h$ and 
$(\Lambda/m_h)^2$ current corrections which grow quadratically.  

The black bursts mark the physical result at the $B_s$ obtained in 
subsection~\ref{subsec:BBs} and the result at the $D_s$ obtained 
using HISQ $c$ and $s$ quarks in~\cite{Donald:2013sra}. 
The mass dependence of our NRQCD results is consistent with a 
value for $R_s$ that grows from our result at the $B_s$ towards the 
result we obtained at the $D_s$ with a relativistic formalism. 
The growth of the NRQCD systematic errors and indeed the fact that 
the $c$ quark in a $D_s$ is not very nonrelativistic mean that 
we cannot accurately extrapolate from results here around the $B_s$ 
to the $D_s$. We can estimate the slope at the $B_s$, however. 
Our results on the coarse lattices, set 4, 
give a linear slope with $m_b/m_h$ for the ratio $R_s$ of 
$0.050(17)$ at a point close to the $b$, where the uncertainty comes 
from NRQCD systematic errors in the current matching.

%
\section{Discussion}
\label{sec:discussion}
%
\begin{figure}[]
\includegraphics[width=0.9\hsize]{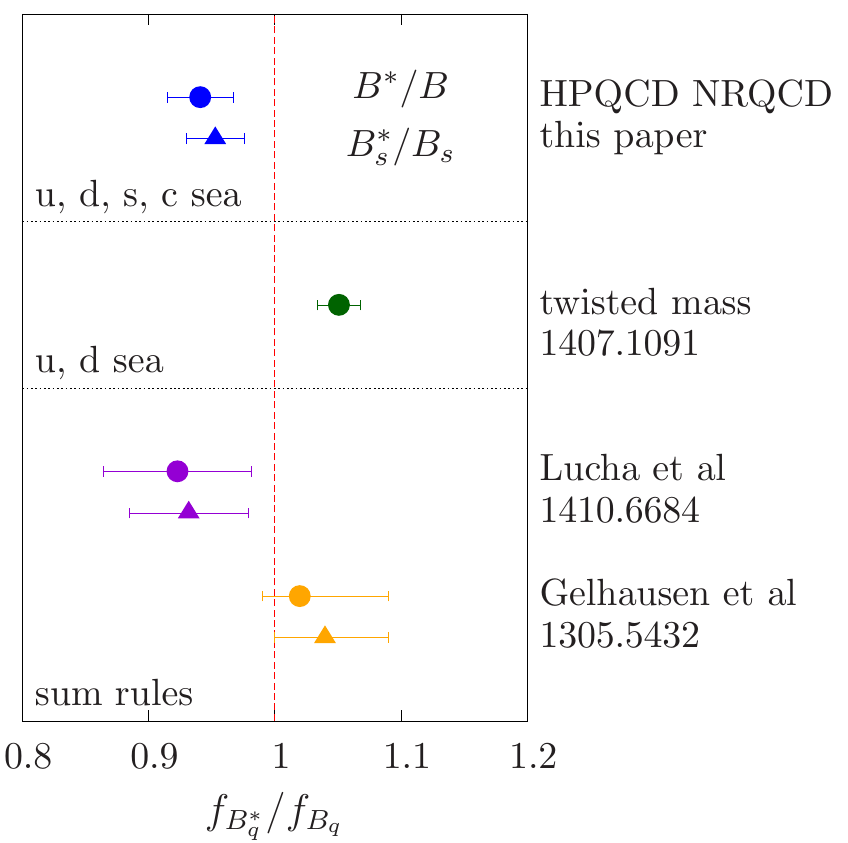}
\caption{A comparison of recent results for the 
ratio of vector to pseudoscalar decay constants for  
$B^*/B$ (filled circles) and $B^*_s/B_s$ (filled triangles). 
The top two results (in blue) are from this paper. 
The filled green triangle is a lattice QCD result using twisted mass 
quarks~\cite{Becirevic:2014kaa} to interpolate 
beween $c$ and the infinite mass limit. 
The lowest two sets of results in purple and orange use 
QCD sum rules~\cite{Lucha:2014nba, Gelhausen:2013wia}.
The red dashed line marks the value 1.0.
}
\label{fig:ratcomp}
\end{figure}

\begin{figure}[]
\includegraphics[width=0.9\hsize]{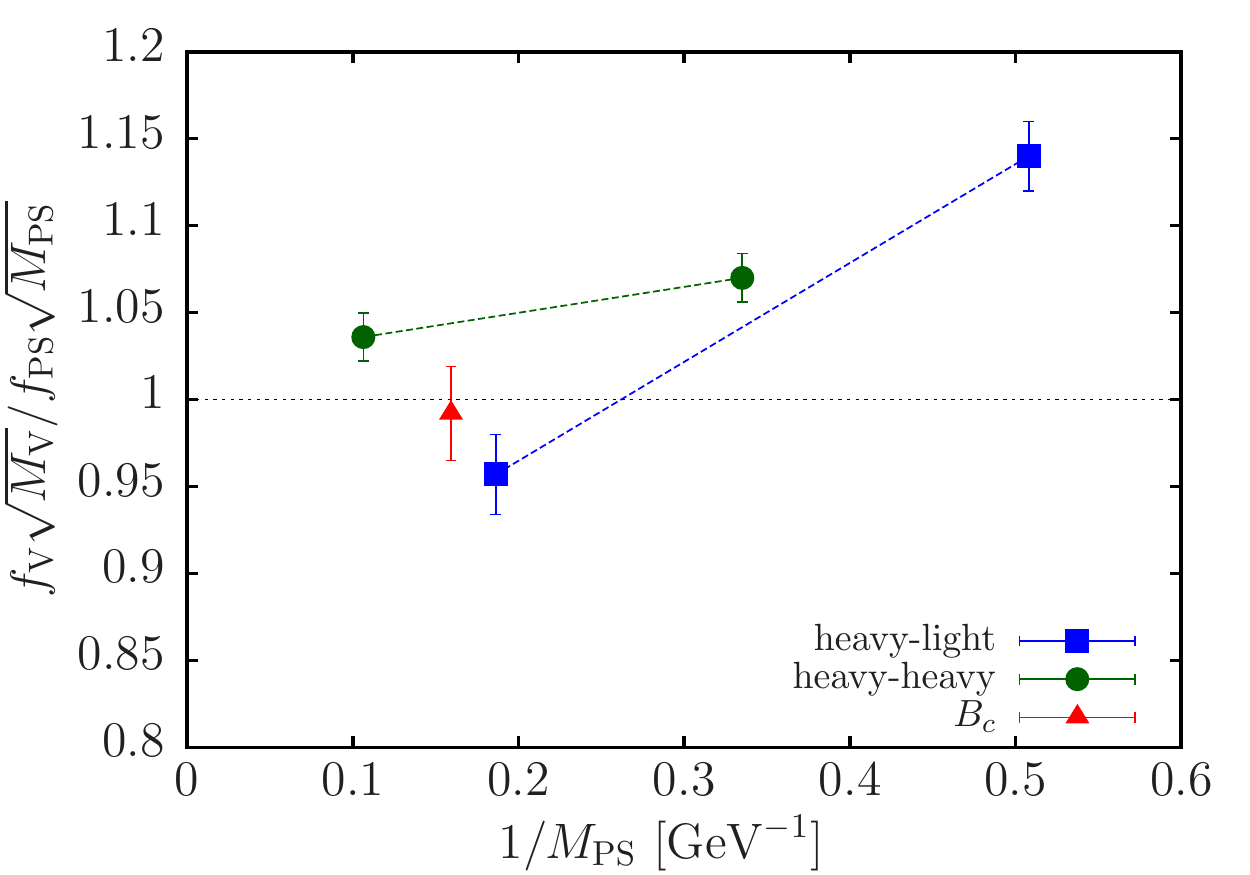}
\caption{The ratio of $f\sqrt{M}$ for vector and pseudoscalar 
mesons with at least one heavy quark, plotted against the inverse of 
the mass for the pseudoscalar meson. Filled blue squares give 
the results obtained here for the $B^*_s/B_s$ along with the results 
from~\cite{Donald:2013sra} for $f_{D^*_s}/f_{D_s}$ multiplied by the 
square root of the ratio of the meson masses from experiment~\cite{pdg}. 
Filled green circles give the results for heavy-heavy mesons using 
experimental values for the vector decay constant obtained from 
the leptonic width~\cite{pdg} and results from lattice QCD using HISQ 
quarks for the pseudoscalar~\cite{newfds,McNeile:2012qf}. 
The filled red triangle gives the result from this paper for the $B^*_c/B_c$ 
ratio. 
}
\label{fig:vpsrat}
\end{figure}

Naively we expect heavy mesons 
with the same valence quark content 
but with vector or pseudoscalar quantum 
numbers to be very similar since spin-dependent
(hyperfine) intereactions that distinguish between 
quark and antiquark having spins parallel or 
anti-parallel are suppressed by the quark mass. 
Such arguments in respect of the meson masses are 
straightforward to make even within the quark model. 
For NRQCD the dominant source of such effects for 
the vector to pseudoscalar meson mass difference 
is the term proportional to $c_4$ in the NRQCD 
Hamiltonian, eq.~(\ref{eq:Hamiltonian})~\cite{Dowdall:2012ab, Dowdallhyp}.

For decay constants the arguments are more subtle which 
is why lattice QCD calculations are important to pin 
down the results. Viewed from the perspective of a nonrelativistic 
effective theory, 
there are three sources for terms 
that affect the ratio of vector to pseudoscalar decay 
constants for heavy mesons: one is the hyperfine term in the Hamiltonian 
as above, the second is the relativistic current correction 
terms ($J^{(1)}$ in eq.~\ref{eq:jpsdef}) and the third is  
matching of the current operator to full QCD. The first two give 
an effect that is proportional to $\Lambda/m_h$ whereas 
the third gives corrections to 1 that are proportional to $\alpha_s$. 
The different dependence of the three effects and the possibilities of 
cancellation between them have given rise to a variety of predictions 
for the ratio of decay constants for vector and 
pseudoscalar heavy-light mesons over the years 
and controversy has surrounded the question of whether the ratio 
is larger or smaller than 1 at the $b$ quark mass. 
Our results here show that the ratio is less than 1 (to $2\sigma$) for $B^*/B$ and 
$B^*_s/B_s$ mesons. In this subsection we set this in the context of earlier 
results.  

A baseline that can be used for heavy-light 
mesons~\cite{Neubert:1992fk} is Heavy 
Quark Effective Theory (HQET) in which the quark 
Lagrangian becomes simple, with no spin dependence, 
in the infinite quark mass limit. 
The matrix elements of the spatial vector and temporal axial 
currents between the vacuum and heavy-light mesons 
become the same in this limit within the effective theory but the renormalisation 
factors that match the currents to full QCD are not 
the same. These have been calculated through 
${\mathcal{O}}(\alpha_s^2)$ in~\cite{Broadhurst:1994se} and 
through ${\mathcal{O}}(\alpha_s^3)$ in~\cite{Bekavac:2009zc} giving, 
to this leading nonrelativistic order and in terms of the $\overline{MS}$ 
coupling~\cite{Bekavac:2009zc}: 
\begin{eqnarray}
\label{eq:threeloop}
\left.\frac{f_B^*}{f_B}\right|_{\mathrm{HQET,LO}} = 1 &-& \frac{2\alpha^{(4)}_s(m_b)}{3\pi} \\
&-& (6.370 + 0.189)\left(\frac{\alpha^{(4)}_s(m_b)}{\pi}\right)^2 \nonumber \\
&-& (77.549 + 6.575) \left( \frac{\alpha^{(4)}_s(m_b)}{\pi} \right)^3 \nonumber \\
&+& \mathcal{O}(\alpha_s^4) . \nonumber
\end{eqnarray}
This is evaluated for $u$, $d$, $s$ and $c$ quarks 
in the sea with the second 
term in the $\alpha_s^2$ and $\alpha_s^3$ coefficients 
taking account of the non-zero mass for the $c$ quark. 
Evaluating the expression 
in eq.~(\ref{eq:threeloop}) gives 0.896~\cite{Bekavac:2009zc}, 
well below 1.0. 

Early calculations added sum-rule 
estimates of $\Lambda/m_h$ hyperfine and current 
corrections to the one-loop piece of eq.~(\ref{eq:threeloop}) 
and obtained a variety of results depending on the relative 
sign of hyperfine and current correction terms. 
In~\cite{Ball:1994uh} it was found that the hyperfine 
and current corrections terms have opposite sign (in
agreement with a subsequent lattice NRQCD study~\cite{sara}) 
and this gave a vector to pseudoscalar decay constant ratio 
for $b$-light mesons of 1.00(4). 
The central value in this result 
would be reduced below 1.0 using the three-loop expression above. 
 
The calculation we give here improves on this approach 
since it is a fully integrated calculation in lattice QCD, 
including dynamics for the $b$ quark from the outset. 
We use an improved NRQCD action for the $b$ quark 
accurate (for heavy-light calculations) 
through $\mathcal{O}(\alpha_s\Lambda/m_h)$
which has been tested on the heavy-light meson 
spectrum~\cite{Dowdall:2012ab} and
from which
we can calculate the matrix elements of current operators 
nonperturbatively. 
The nonrelativistic current, combining the leading term and 
first, $\Lambda/m_b$, 
relativistic correction is matched to full QCD 
and the $\mathcal{O}(\alpha_s)$ matching correction is found to be very small. 

Our results, as described in Subsection~\ref{subsec:BBs}, 
show that $f_{B^*_s}/f_{B_s}$ and $f_{B^*}/f_B$ are 
about 5.0(2.5)\% below 1, and the ratio for the $B^*/B$ is 
1.3(1.4)\% below that of the $B^*_s/B_s$. 
Our values are compared to results from two recent QCD sum-rule 
analyses~\cite{Lucha:2014nba, Gelhausen:2013wia} 
in Figure~\ref{fig:ratcomp}. Although there is some tension,
those results are consistent with 
each other and with our numbers here. 
All the results show the same tendency for the ratio for 
$B^*/B$ to be slightly smaller than for $B^*_s/B_s$, 
although the difference is not significant in any of the cases. 

We also compare to a recent lattice QCD result~\cite{Becirevic:2014kaa} 
which used the twisted-mass formalism for both heavy 
and light quarks on gluon field configurations that included 
the effect of $u/d$ quarks (only) in the sea. 
The twisted-mass value is obtained 
from results calculated for heavy quark masses around the 
$c$ quark mass and above. 
An interpolation between those results and the infinite mass 
limit is performed to reach the $b$, using the first
two-loops of the three-loop formula of eq.~(\ref{eq:threeloop}) to rescale results 
so that 1.0 (up to higher-order corrections) is obtained in the infinite mass limit.  
The value 
quoted for $f_{B^*}/f_B$ is 1.051(17) and this 
disagrees with our value by more than 3 (combined)
standard deviations. 
It is not clear that results using only $u/d$ quarks 
in the sea will necessarily agree with those, like ours, 
that include a full complement of sea quarks. 
This may be a case where the `quenching' of the $s$ 
quark produces a visible effect. 
A more likely source of difference is probably the 
interpolation in~\cite{Becirevic:2014kaa} between 
the charm mass and the infinite mass limits. Such 
an interpolation requires evaluating the formula 
of eq.~(\ref{eq:threeloop}) using $\alpha_s$ at 
a scale much lower than $m_b$ where the relatively 
large coefficients
make that problematic. 

It is also interesting to compare results for 
the ratio of vector to pseudoscalar decay constants 
between heavy-heavy mesons and heavy-light mesons. 
The decay constant of vector heavyonium mesons can 
be determined from their experimental decay rate to 
leptons:
\begin{equation}
\label{eq:leptdecay}
\Gamma(v_h \rightarrow e^+e^-) = \frac{4\pi}{3}\alpha_{\mathrm{QED}}^2e_h^2\frac{f_v^2}{m_v} .
\end{equation}
The decay constants can also be calculated in lattice 
QCD~\cite{Becirevic:2012dc,Donald:2012ga, Colquhoun:2014ica} and 
good agreement with experiment is found. 
Since heavyonium pseudoscalar mesons do not annihilate to a single 
particle, there is no direct experimental determination of 
the decay constant. Again, however, the decay constants can 
be accurately determined in lattice QCD~\cite{newfds, McNeile:2012qf}. 

Figure~\ref{fig:vpsrat} shows the ratio of vector to pseudoscalar 
decay constants (multiplied by the square root of the ratio 
of the masses) for $(J/\psi)/\eta_c$, $\Upsilon/\eta_b$, $B^*_s/B_s$ 
and $D_s^*/D_s$ plotted against the inverse of the corresponding 
pseudoscalar meson mass. For the $J/\psi$ and $\Upsilon$ decay 
constants we use the values determined from the experimental 
annihilation rates~\cite{pdg} and eq.~(\ref{eq:leptdecay}). 
These are 0.407(5) GeV and 0.689(5) GeV respectively. 
From full lattice QCD the $\eta_c$ decay constant 
is 0.3947(24) GeV~\cite{newfds} and 
the $\eta_b$ decay constant is 0.667(6) GeV~\cite{McNeile:2012qf}. 
The $D_s^*/D_s$ decay constant ratio is taken from~\cite{Donald:2013sra}. 
We see that the behaviour for heavyonium and heavy-light mesons is 
similar but the slope is larger for heavy-light mesons. 

For heavyonium mesons, similar considerations apply to the 
decay constant ratio as discussed above for heavy-light mesons.
A baseline might be considered a simple spin-independent potential 
model in which the decay constant can be related to 
the `wavefunction-at-the-origin'. However there are significant 
QCD radiative corrections to match $\psi(0)$ to the decay constant 
in both the vector (see, for example~\cite{Barbieri:1975jd}) 
and pseudoscalar~\cite{Braaten:1995ej} cases, and these need 
to be included. 
Going beyond this requires the inclusion of spin-dependent 
terms in the Hamiltonian and relativistic corrections to the 
leading-order current. These are taken care of in a lattice QCD 
calculation, either explicitly when using a nonrelativistic 
formalism such as NRQCD~\cite{Colquhoun:2014ica} or 
implicitly included when using a relativistic formalism such 
as HISQ~\cite{McNeile:2012qf}.   

Here we have calculated the decay constant of both the $B_c$ 
and the $B^*_c$, using NRQCD $b$ quarks and HISQ $c$ quarks and 
working through first-order in the QCD matching and relativistic 
spin-dependent corrections to the NRQCD Hamiltonian and the 
currents. 
Our result for the $B_c$ decay constant agrees 
well with that obtained previously using the relativistic 
HISQ formalism for both $b$ and $c$ quarks~\cite{McNeile:2012qf}, 
adding confidence to our analysis of systematic errors 
in both the nonrelativistic and relativistic approach. Here we also 
calculate the ratio of decay constants for the $B_c^*$ and $B_c$, 
for the first time from lattice QCD.

$B_c$ and $B^*_c$ decay
constants have also been calculated 
within a potential-model approach, including QCD 
radiative corrections. See~\cite{Kiselev:2003uk} for a discussion. 
Results are in reasonable agreement with ours, 
but with a larger uncertainty because the approach has less control 
of systematic errors.  

We find a value for $f_{B^*_c}/f_{B_c}$ which is larger 
than that of $f_{B^*_s}/f_{B_s}$, indicating that the 
internal structure of the $B_c$ is somewhat different from 
that of a typical heavy-light meson. Figure~\ref{fig:vpsrat} 
shows this clearly. When the decay constant ratio is 
plotted for the $B^*_c/B_c$ it lies very neatly between 
the heavy-heavy line and the heavy-light line. 
 
%
\section{Conclusions}
\label{sec:conclusions}
%
\begin{figure*}[]
\includegraphics[width=0.8\hsize]{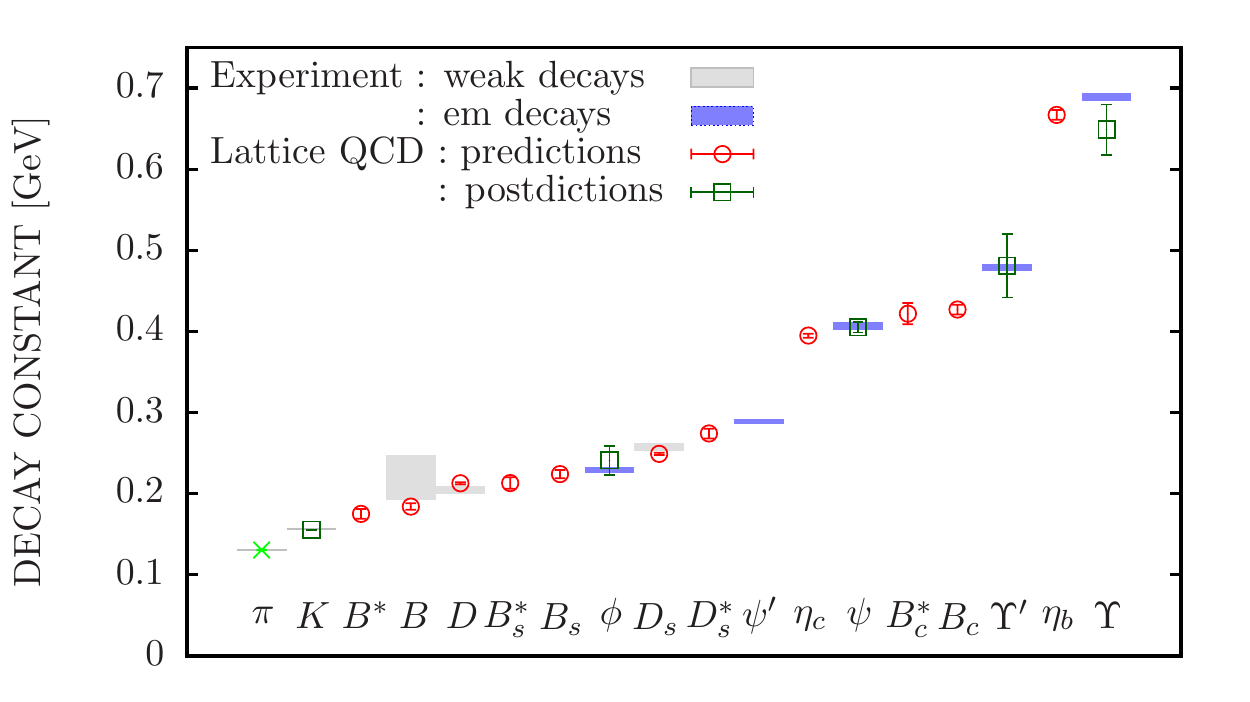}
\caption{A summary of values for decay constants of mesons that are narrow 
and so well-characterised in experiment. Experimental values are given 
as blue or grey bands and are taken from average weak or electromagnetic 
annihilation rates~\cite{pdg} using, for weak decays, average values of 
the appropriate CKM matrix element. 
For full lattice QCD results, green open squares (postdictions) or red 
open circles (predictions), we 
take world's best values. The lattice result for $f_{\pi^+}$ is marked with 
a cross to indicate that it is used to set the scale in some analyses (although 
not here). The result for the $K^+$ is from~\cite{Dowdall:2013rya}, 
the $B^+$ and $B_s$ from~\cite{Dowdall:2013tga}, 
the $D^+$ and $D_s$ from~\cite{Bazavov:2014wgs},  
the $\phi$ from~\cite{Donald:2013pea}, 
the $D^*_s$ from~\cite{Donald:2013sra},
the $\eta_c$ from~\cite{newfds}, 
the $J/\psi$ from~\cite{Donald:2012ga}, 
the $B_c$ and $\eta_b$ from~\cite{McNeile:2012qf}, 
the $\Upsilon$ and $\Upsilon^{\prime}$ from~\cite{Colquhoun:2014ica} 
and the $B^*$, $B^*_s$ and $B^*_c$ from this paper.
}
\label{fig:decaygold}
\end{figure*}

Decay constants, which parameterise the amplitude for a 
meson to annihilate to a single particle, are as much a part 
of a meson's `fingerprint' as its mass. 
They are often harder to determine, however, and some cannot be 
accessed directly through an experimental decay rate. 
The overall picture of meson decay constants gives information 
about how the internal structure of mesons changes for 
different quark configurations as a result of QCD interactions. 
To obtain this picture in sufficient detail, for example even 
to put the decay constants into an order, requires 
calculations in full lattice QCD, since only then can 
we reliably quantify the systematic errors. 

Here we have expanded range of decay constant calculations 
from full lattice QCD 
to include vector heavy-light mesons. 
Our results for the ratio of vector to pseudoscalar decay 
constants are: 
\begin{eqnarray}
\frac{f_{B^*}}{f_B} &=& 0.941(26) \\
\frac{f_{B^*_s}}{f_{B_s}} &=& 0.953(23) \nonumber \\
\frac{f_{B^*_c}}{f_{B_c}} &=& 0.988(27) \nonumber.
\end{eqnarray} 
Thus
\begin{itemize}
\item The vector decay constant is smaller than the pseudoscalar decay 
constant for $b$-light mesons, at the $2\sigma$ level for $B^*/B$ and $B^*_s/B_s$. 
This is in contrast to results for $c$-light mesons where the
vector has a larger decay constant than the pseudoscalar.
\item The ratio of vector to pseudoscalar decay constants shows 
an ordering so that $f_{B^*_c}/f_{B_c} > f_{B^*_s}/f_{B_s} > f_{B^*}/f_B$. 
When correlations between the uncertainties are taken into 
account using ratios, the first of these relationships 
has $3\sigma$ significance, the 
second $1\sigma$ (see eqs.~(\ref{eq:rcrsres}) and (\ref{eq:rlrsres})). 
\end{itemize}

Using our earlier world's best results for $f_B$ 
(0.186(4) GeV, isospin-averaged), $f_{B_s}$ (0.224(5) GeV)~\cite{Dowdall:2013tga}
and $f_{B_c}$ (0.427(6) GeV)~\cite{McNeile:2012qf} we derive values for the 
vector decay constants: 
\begin{eqnarray}
f_{B^*} &=& 0.175(6) \,\mathrm{GeV} \\
f_{B^*_s} &=& 0.213(7)\, \mathrm{GeV} \nonumber \\
f_{B^*_c} &=& 0.422(13)\, \mathrm{GeV} \nonumber.
\end{eqnarray} 

Finally, in Figure~\ref{fig:decaygold} we 
give a `spectrum' plot for the decay constants 
of 15 gold-plated mesons from lattice QCD, including 
the new results from this paper. It illustrates 
the coverage and predictive power of lattice 
QCD calculations. 
The decay constants are ordered by value, something that 
is only possible with sufficiently accurate results. 
The range of values is much smaller than that for meson 
masses and the ordering of values is not as obvious 
because the quark masses do not have the same impact 
on the decay constants as they do on the meson masses. 
The plot therefore shows up some interesting features in 
the ordering, for example that the $K$ and $B^*$ mesons 
have such similar values and that the $\phi$ meson appears so 
far up the list. 
We see that the decay constants for vector-pseudoscalar 
pairs are close together everywhere, closer than for the 
pairings in which an $s$ quark is substituted for a light 
quark in a meson, for example.  

Future work will improve the accuracy of lattice QCD results 
for the vector-onium states such as the 
$\phi$ (not strictly gold-plated)~\cite{Chakraborty:2014zma} 
and the $\psi^{\prime}$~\cite{Galloway:2014tta}, both of which can be 
determined accurately from experiment. The issues there are mainly 
from lattice QCD statistical errors. For $b$-light meson decay constants the 
dominant source of uncertainty, as we have seen, is from 
systematic errors in NRQCD such as current renormalisation factors. 
Work is underway to reduce these further using techniques based on 
current-current correlator methods~\cite{Colquhoun:2014ica,Koponen:2010jy}.

%
\section*{Acknowledgements}
\label{sec:acknowledgements}
%
We are grateful to the MILC collaboration for the use of their 
gauge configurations, to R. Horgan. C.Monahan and J. Shigemitsu for 
calculating the pieces needed for the current renormalisation 
used here, and to B. Chakraborty, A. Grozin, K. Hornbostel, F. Sanfilippo 
and S. Simula for useful discussions. 
The results described here were obtained using the Darwin Supercomputer 
of the University of Cambridge High Performance 
Computing Service as part of the DiRAC facility jointly
funded by STFC, the Large Facilities Capital Fund of BIS 
and the Universities of Cambridge and Glasgow. 
This work was funded by STFC, NSF, the Royal Society and the Wolfson Foundation.

\bibliography{hl_bib}

\end{document}